\documentclass[a4paper,11pt]{article}
\usepackage{jheppub}
\usepackage{amsmath,amssymb,graphicx,float,slashed,xcolor,multicol}
\usepackage{enumerate}
\usepackage{caption,subcaption}
\usepackage{float}
\usepackage{multirow}
\usepackage[normalem]{ulem}
\usepackage{tablefootnote}
\title{Radiative Decays of the Higgs Boson to a Pair of Fermions}
\author[a,b]{Tao Han}
\author[a]{and Xing Wang}
\affiliation[a]{Pittsburgh Particle physics, Astrophysics, and Cosmology Center, \\
Department of Physics and Astronomy, University of Pittsburgh, \\
3941 O'Hara St., Pittsburgh, PA 15260, USA}
\affiliation[b]{Department of Physics, Tsinghua University, and Collaborative Innovation Center of Quantum Matter, Beijing, 100086, China}
\emailAdd{than@pitt.edu}
\emailAdd{xiw77@pitt.edu}
\abstract{
 We revisit the radiative decays of the Higgs boson to a fermion pair $h\rightarrow f\bar{f}\gamma$ where $f$ denotes a fermion in the Standard Model (SM). We include the chirality-flipping diagrams via the Yukawa couplings at the order $\mathcal{O}(y_f^2 \alpha)$, the chirality-conserving contributions via the top-quark loops of the order $\mathcal{O}(y_t^2 \alpha^3)$, and the electroweak loops at the order $\mathcal{O}(\alpha^4)$. The QED correction is about $Q_f^2\times {\cal O}(1\%)$ and contributes to the running of fermion masses at a similar level, which should be taken into account for future precision Higgs physics.
The chirality-conserving electroweak-loop processes are interesting from the observational point of view. First, the branching fraction of the radiative decay $h \to \mu^+\mu^- \gamma$ is about a half of that of $h \to \mu^+\mu^-$, and that of $h \to e^+ e^- \gamma$ is more than four orders of magnitude larger than that of $h \to e^+ e^-$, both of which reach about $10^{-4}$. The branching fraction of $h \to \tau^+\tau^- \gamma$ is of the order $10^{-3}$. 
All the leptonic radiative decays are potentially observable at the LHC Run 2 or the HL-LHC.
The kinematic distributions for the photon energy or the fermion pair invariant mass provide non-ambiguous discrimination for the underlying mechanisms of the Higgs radiative decay. We also study the process $h \to c\bar c \gamma$ and evaluate the observability at the LHC. We find it potentially comparable to the other related studies and better than the $h \to J/\psi\ \gamma$ channel in constraining the charm-Yukawa coupling.
}
\preprint{
\begin{flushright}
PITT-PACC-1703
\end{flushright}
}

\begin{document}
\maketitle

\section{Introduction}

The discovery of the Higgs boson at the CERN Large Hadron Collider (LHC) has set a milestone in particle physics \cite{Aad:2012tfa, Chatrchyan:2012xdj}. All the studies indicate that it is consistent with the Standard Model (SM) Higgs boson \cite{Khachatryan:2016vau}. However, there are only a handful channels observed at the LHC and accuracies on the branching fraction measurements, even assuming it is the SM Higgs boson, are still no better than about $10\%$. 
There are compelling motivations that the SM needs to be extended, including the particle dark matter, the origin of the neutrino mass, and perhaps the most puzzling related to the electroweak scale, the ``naturalness'' for the Higgs boson mass in the SM. Therefore, more detailed studies regarding the properties of the Higgs boson are necessary to test the SM and to look for possible new physics beyond the Standard Model.

With a large amount of data being accumulated at the LHC Run-2 and the higher luminosity expectation 
of 3 ab$^{-1}$ (HL-LHC), one would expect to produce a large sample, eventually reaching about 50 pb$\times$ 3 ab$^{-1}\approx 150$ million Higgs bosons. As thus, searching for rare decays of the Higgs boson becomes feasible and thus increasingly important to test the Higgs sector in the Standard Model and to seek for new physics beyond the SM. One important feature of the Higgs boson predicted by the SM is that the Higgs-fermion interaction strength, the Yukawa coupling, is proportional to the fermion mass. So far, ATLAS \cite{ATLAS:2014lua, Aad:2014xzb, Aad:2015vsa} and CMS \cite{Khachatryan:2014qaa, Chatrchyan:2013zna, Chatrchyan:2014nva} have only been able to measure the Higgs couplings to the third-generation fermions ($\tau$ and $b$ directly and $t$ indirectly). 
The Higgs rare decays to a light fermion pair are usually very difficult to observe because of the suppression by the small Yukawa couplings. For instance, the branching fraction of $h\rightarrow e^+e^-$ is $\mathcal{O}(10^{-8})$, and thus hopeless to detect this decay channel at colliders. 

In this paper, we study other rare decay channels: the Higgs radiative decay to a fermion pair $h\rightarrow f\bar{f}\gamma$. Firstly, this decay channel receives contribution that is proportional to the Higgs-fermion interaction strength, which may provide a complementary way to measure certain Yukawa couplings. Secondly, as it also receives contributions from electroweak (EW) one-loop diagrams \cite{Abbasabadi:1996ze}, this channel is not necessarily governed by the Yukawa coupling for light fermions, leading to violation of the Yukawa scaling.
Due to this enhancement, the Higgs transitions to light fermions may be observable via the radiative decays despite the smallness of fermion masses. The searches for those Higgs decays are not only to test the consistency of the SM, but also to seek for potential new physics in either the Yukawa or the electroweak sector \cite{Li:1998rp, Arhrib:2014pva, Hu:2014eia, Belanger:2014roa}. 
We present our systematical treatment to such channels from the observational points of view at the LHC. We lay out the kinematical features for the leptonic channels $h\to \mu^+\mu^- \gamma,  e^+e^- \gamma$ near $\gamma^*, Z$ poles and the interplay in between, and propose new cuts based on the kinematical features to optimize the on-going searches. We also motivate a new search for $h\to \tau^+ \tau^- \gamma$ which should be within the scope of observability for a Higgs rare decay at the LHC. Furthermore, we propose another new channel $h \to c \bar c \gamma$ to be searched for at the LHC, which could complement the existing proposals on probing the charm-quark Yukawa coupling at the LHC. In the due course, we point out the numerical significance of the QED running mass concerning the future precision Higgs measurements of the Yukawa couplings.

The rest of the paper is organized as follows. We present the full one-loop electroweak corrections to the decay $h\to f\bar f$ in Sec.~\ref{sec:cal} and show the kinematical features by some differential distributions. 
We then discuss the observability of the leptonic channels at the LHC in Sec.~\ref{sec:LHCll}. We finally study the difficult channel $h\to c\bar c \gamma$ in Sec.~\ref{sec:LHCcc}. We summarize our results in Sec.~\ref{sec:sum}. 


\section{$h\to f\bar f (\gamma)$ at One-Loop}
\label{sec:cal}

\begin{figure}
\centering
\begin{subfigure}[b]{0.23\textwidth}
\includegraphics[width=\textwidth]{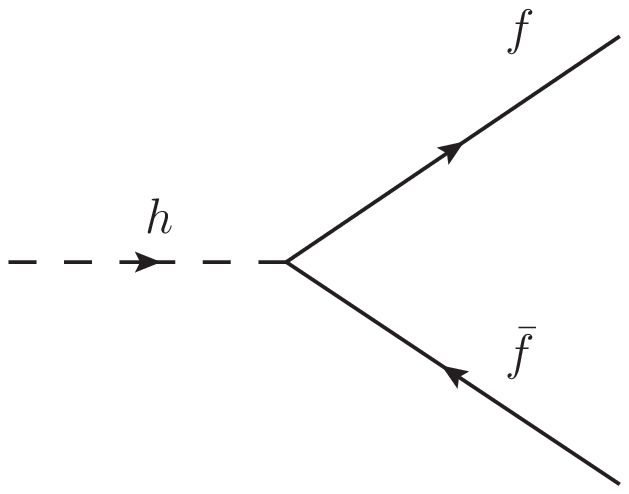}
\caption{}
\label{feyn:born}
\end{subfigure}
\begin{subfigure}[b]{0.23\textwidth}
\includegraphics[width=\textwidth]{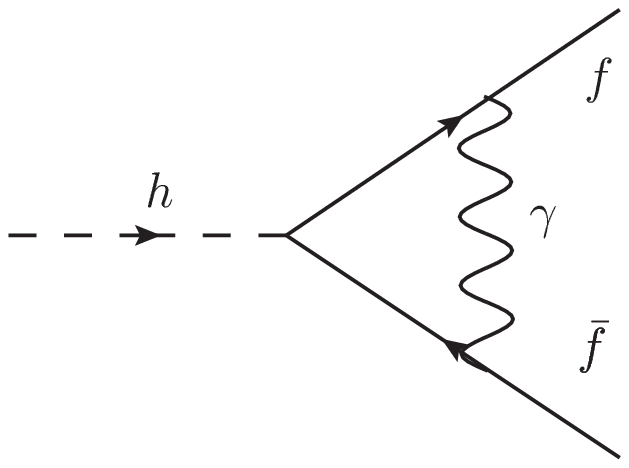}
\caption{}
\label{feyn:qed1}
\end{subfigure}
\begin{subfigure}[b]{0.23\textwidth}
\includegraphics[width=\textwidth]{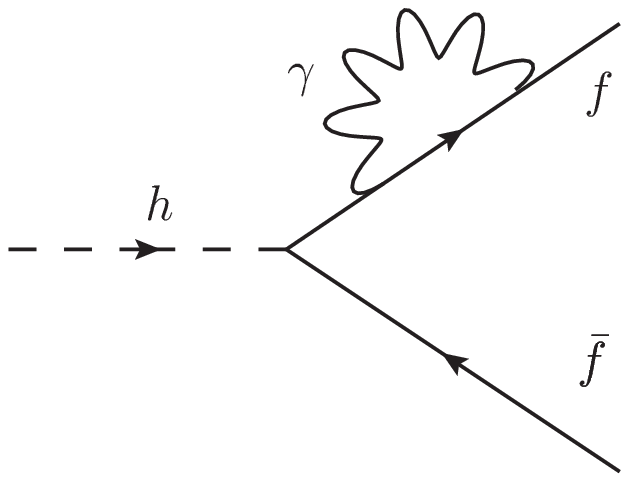}
\caption{}
\label{feyn:qed2}
\end{subfigure}
\begin{subfigure}[b]{0.25\textwidth}
\includegraphics[width=\textwidth]{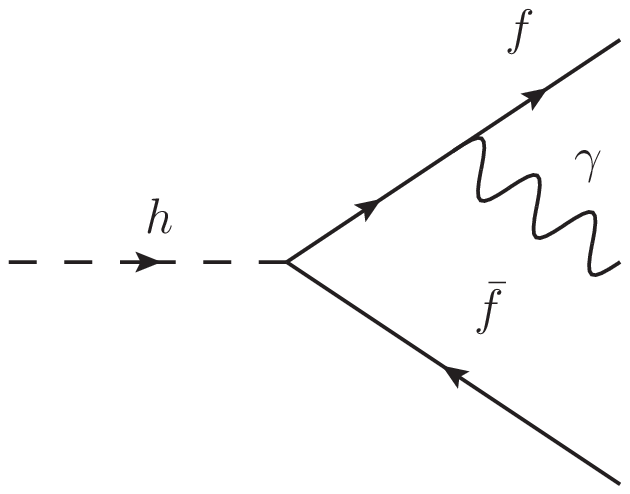}
\caption{}
\label{feyn:qed3}
\end{subfigure} \\
\begin{subfigure}[b]{0.23\textwidth}
	\includegraphics[width=\textwidth]{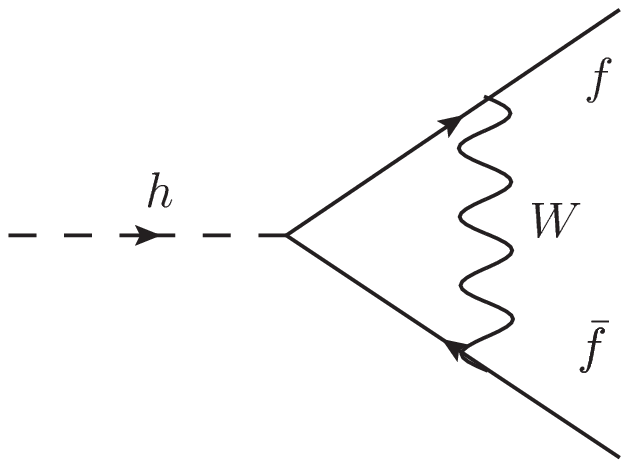}
	\caption{}
	\label{feyn:qed4}
\end{subfigure}
\begin{subfigure}[b]{0.23\textwidth}
	\includegraphics[width=\textwidth]{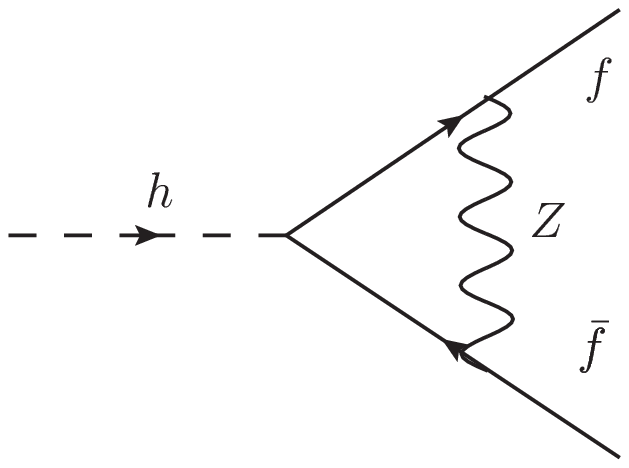}
	\caption{}
	\label{feyn:qed5}
\end{subfigure}
\begin{subfigure}[b]{0.25\textwidth}
	\includegraphics[width=\textwidth]{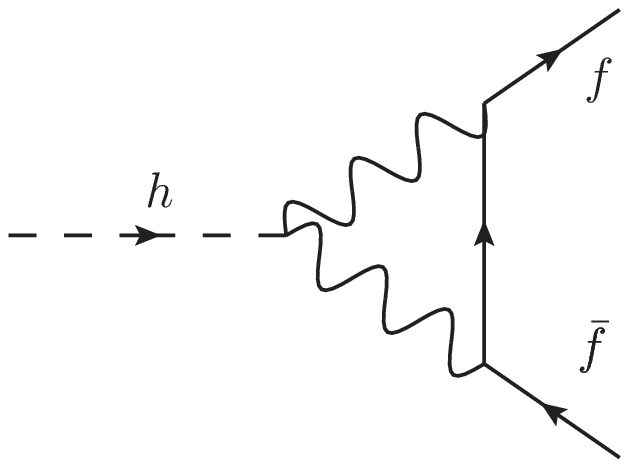}
	\caption{}
	\label{feyn:qed6}
\end{subfigure}
\caption{Representative Feynman diagrams of $h\rightarrow f\bar{f}$ and its EW radiative corrections up to $\mathcal{O}(y_f^2\alpha)$. }
\label{feyn:qed}
\end{figure}
%


It is well known that the tree-level decay width for $h\to f \bar f$ as shown in Fig.~\ref{feyn:born} is
\begin{equation}
\Gamma_{h\rightarrow f \bar{f}}^0 = \frac{y_f^2 N_c}{16\pi}\ m_h\ \beta_f^3,\qquad \beta_f = \sqrt{1-{4m_f^2 \over m_h^2}}.
\label{eq:gm}
\end{equation}
where, in the SM, the Yukawa coupling is $y_f = \sqrt 2\ m_f/v$, and the color factor $N_c=3\ (1)$ for a color triplet (singlet) fermion. Quantum chromodynamics (QCD) corrections to the decay of the Higgs to a quark pair have been known up to N$^4$LO at $\mathcal{O}(\alpha_s^4)$ \cite{Djouadi:2005gi,Baikov:2005rw, Davies:2017xsp}. To serve as a comparison with the current work, we write the expression as
\begin{equation}
\Gamma_{\rm NLO\ QCD} = \Gamma^0 \left( 1+ C_F \frac{\bar \alpha_s}{\pi}\ {17\over 4} 
+ \mathcal{O}(\alpha^2_s)\right), \quad
\Gamma^0 = \frac{N_c}{8\pi}\ m_h\ {\bar m_f^2 \over v^2}\ \beta_f^3,
\label{eq:qcd}
\end{equation}
where $\bar\alpha_s^2$ and $\bar m_f^2$ are the renormalized QCD running coupling and quark mass, respectively, to the scale $m_h^2$ in the $\overline{\rm MS}$ subtraction scheme, and the color factor $C_F=(N_c^2-1)/2N_c=4/3$. The most significant effect is due to the running of the quark mass from $\mu_0=m_f$ to $\mu=m_h$ \cite{Braaten:1980yq,Sakai:1980fa,Inami:1980qp,Gorishnii:1983cu,Drees:1990dq}. For the sake of illustration and comparison, we only give the one-loop QCD running mass expression as 
\begin{align}
\bar{m}(\mu) = \bar{m}(\mu_0) \left(\frac{\bar{\alpha}_s(\mu)}{\bar{\alpha}_s(\mu_0)}\right)^{\gamma_0\over b_0}= \ \bar{m}(\mu_0)\left(1+{b_0 \over 4\pi} \bar{\alpha}_s(\mu_0)\ln{\mu^2\over \mu_0^2} \right)^{-{\gamma_0\over b_0}} 
\label{eq:qcd_running_mass}
\end{align}
where $\gamma_0 = 4$ and $b_0 = 11 - 2n_f/3$ in QCD.


\subsection{$\mathcal{O}(y_f^2\alpha)$ corrections}
\label{sec-QED}

\begin{table}[tb]
	\begin{center}
		\renewcommand{\arraystretch}{1.5}
		\begin{tabular}{|c|c|c|c|c|c|}
			\hline
			Fermion & $\bar{m}_f (m_f)$ & $\delta \bar{m}^{\rm QCD}_f$ & $\delta \bar{m}^{\rm QED}_f$ & $\bar{m}_f(m_h)$ & $\Gamma^0_{h\rightarrow f\bar{f}}$\\
			& [GeV] & [GeV] & [MeV] & [GeV] & [keV]\\
			\hline\hline
			$b$ & 4.18 & $-1.39$ & $-5.72\;(1\%)$ & 2.78 & 1900\\
			\hline
			$c$ & 1.27 & $-0.657$  & $-9.33\;(0.7\%)$ & 0.604 & 89.7\\
			\hline\hline
			$\tau$ & 1.78 & - & $-27.2\;(0.4\%)$ & 1.75 & 251\\
			\hline
			$\mu$ & 0.106 & - & $-4.05\;(0.2\%)$ & 0.102 & 0.852\\
			\hline
			$e$ & $0.511 \times 10^{-3}$ & - & $-2.20\times 10^{-2}\;(0.1\%)$ & $0.489\times 10^{-3}$ & $1.96\times 10^{-5}$\\
			\hline
		\end{tabular}
	\end{center}
	\caption{The $\overline{\rm MS}$ running masses with N$^4$LO QCD and NLO QED corrections. The last column is the LO width with the running Yukawa coupling effect. The relative size of the differences between the QED resummed running mass in Eq.~(\ref{eq:qcd_running_mass}) and its $\mathcal{O}(\alpha)$  approximation are given in the parentheses.}
	\label{tab_mass}
\end{table}

Similar to the above, QED corrections to the Higgs radiative decay at $\mathcal{O}(y_f^2\alpha)$, depicted 
in Figs.~\ref{feyn:qed1}$-$\ref{feyn:qed3}, have the same form except for the color factor and the electric charge of the fermions \cite{Kataev:1997cq},
\begin{equation}
\Gamma_{\rm NLO\ QED} = \Gamma^0 \left(1+ Q_f^2\frac{\bar \alpha}{\pi}\ \frac{17}{4} + \mathcal{O}(\alpha^2)\right).
\end{equation}
Therefore, the QED corrections to the partial width at the next-to-leading order (NLO) contribute about $Q_f^2\times \mathcal{O}(1\%)$ to the Higgs partial width to a fermion pair.
Analogous to QCD, we should also take into account the effect of QED running mass, which can be calculated using Eq.~(\ref{eq:qcd_running_mass}) with $\gamma_0 = 3Q_f^2$ and $b_0 = - 4\sum_f Q_f^2/3$ in QED. This 1-loop running from $m_f$ to $m_h$ will change the fermion mass by about $4\%$ for the electron and about $0.1\%\ (0.8\%)$ for the $b$-quark ($c$-quark), comparable to the fix-oder QED correction as above. The running mass effect from N$^4$LO QCD \cite{Chetyrkin:1997dh, Vermaseren:1997fq} and NLO QED are summarized in Table \ref{tab_mass}. The difference between the QED resummed running mass in Eq.~(\ref{eq:qcd_running_mass}) and its $\mathcal{O}(\alpha)$ expansion is relatively small due to the weakly-coupled nature of QED, and contributes to the NLO QED corrections at percentage level, as shown in the parentheses in Table \ref{tab_mass}.
The entries in the last column of Table \ref{tab_mass} are evaluated with the running Yukawa coupling effects, using the LO partial width formula of Eq.~(\ref{eq:gm}). 
We note that the full SM prediction for the Higgs total width is 4.1 MeV \cite{deFlorian:2016spz}.

The complete EW corrections to $h\rightarrow f\bar{f}$ partial width at $\mathcal{O}(y_f^2\alpha)$ is
\begin{equation}
\delta \Gamma_{\rm EW} = \Gamma^0 \left( \frac{2\delta m_f^{\rm QED}}{\bar{m}_f} + Q_f^2\frac{\bar \alpha}{\pi}\ \frac{17}{4} + \Delta_{\rm weak} + \mathcal{O}(\alpha^2)\right),
\label{eq:fullew}
\end{equation}
where $\delta m_f^{\rm QED} = \bar{m}(m_h)- \bar{m}(m_f)$ as listed in Table~\ref{tab_mass}, and 
$\Delta_{\rm weak}$ follows the on-shell definition in \cite{Kniehl:1993ay}. The two terms of QED are for mass and vertex corrections and they have opposite signs.
The 1-loop EW diagrams as shown in Figs.~\ref{feyn:qed1}$-$\ref{feyn:qed6} are all proportional to $m_f$, and thus we will refer this section as ``Yukawa corrections''. We also refer the exclusive real photon emission represented by Fig.~\ref{feyn:qed3} as ``QED radiation'' in later sections.
EW corrections with higher-order loops up to $\mathcal{O}(\alpha\alpha_s)$ \cite{Mihaila:2015lwa} and $\mathcal{O}(\alpha_s^2 G_F m_t^2)$ \cite{Chetyrkin:1996ke, Chetyrkin:1996wr} have also been calculated, that we will not include in the current study.

 As the precision of the Higgs measurements improves in the future, it will become necessary to take these corrections into account. In particular, the projected precision of the $hb\bar{b}$ coupling determination was estimated to be $0.3\%$ at the International Linear Collider \cite{Peskin:2013xra}. 

\subsection{$\mathcal{O}(y_t^2\alpha^3, \alpha^4)$ corrections}
\label{sec-EW}

\begin{figure}[tb]
\centering
\begin{subfigure}[b]{0.23\textwidth}
\includegraphics[width=\textwidth]{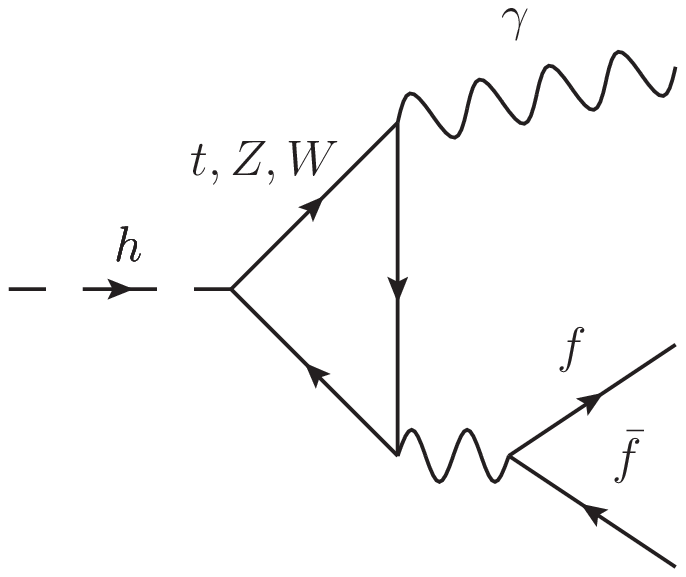}
\caption{}
\label{feyn:ew1}
\end{subfigure}
\begin{subfigure}[b]{0.23\textwidth}
\includegraphics[width=\textwidth]{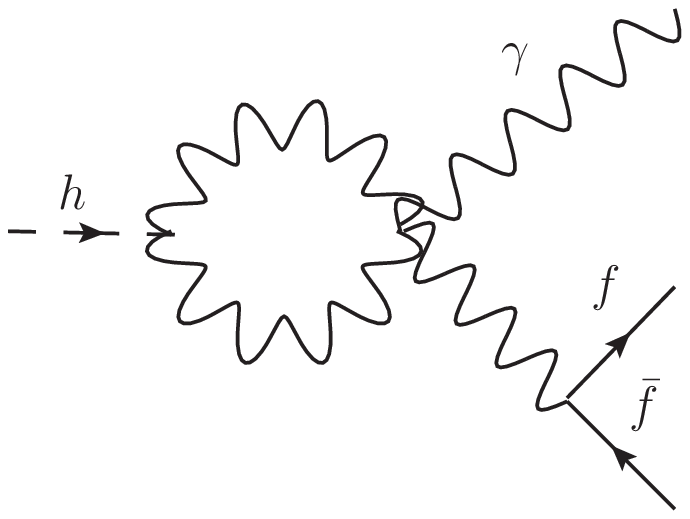}
\caption{}
\label{feyn:ew2}
\end{subfigure}
\begin{subfigure}[b]{0.23\textwidth}
\includegraphics[width=\textwidth]{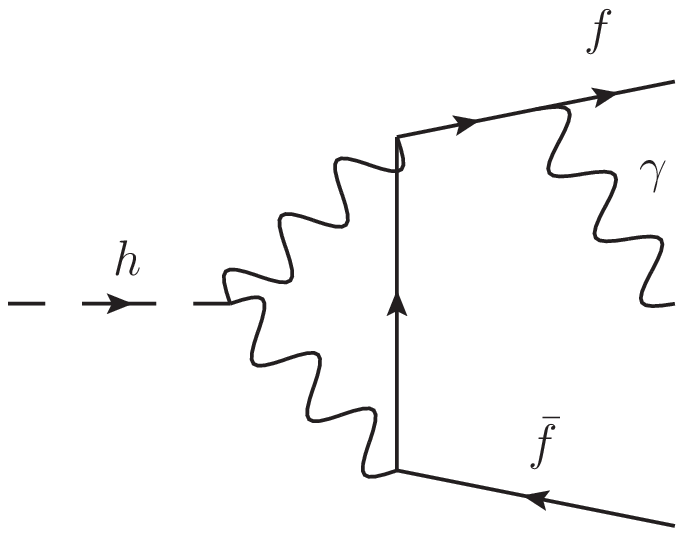}
\caption{}
\label{feyn:ew3}
\end{subfigure}
\begin{subfigure}[b]{0.23\textwidth}
\includegraphics[width=\textwidth]{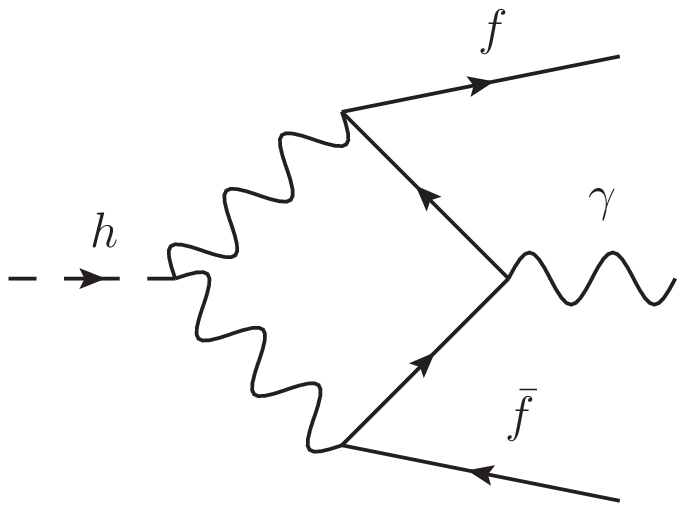}
\caption{}
\label{feyn:ew4}
\end{subfigure}
\\
\begin{subfigure}[b]{0.23\textwidth}
\includegraphics[width=\textwidth]{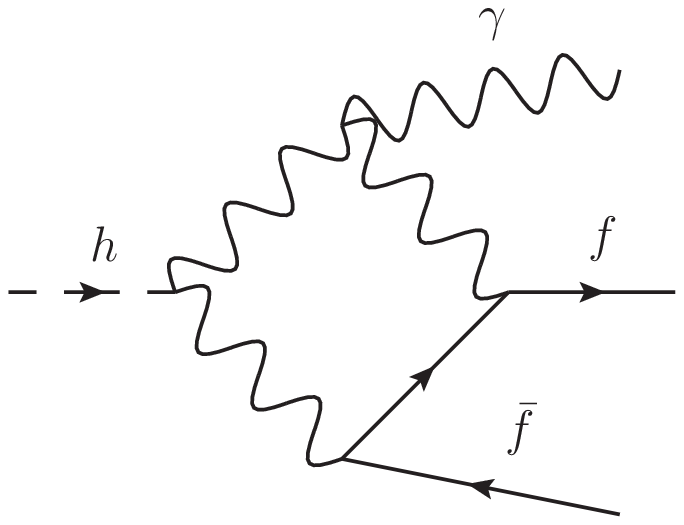}
\caption{}
\label{feyn:ew5}
\end{subfigure}
\begin{subfigure}[b]{0.23\textwidth}
	\includegraphics[width=\textwidth]{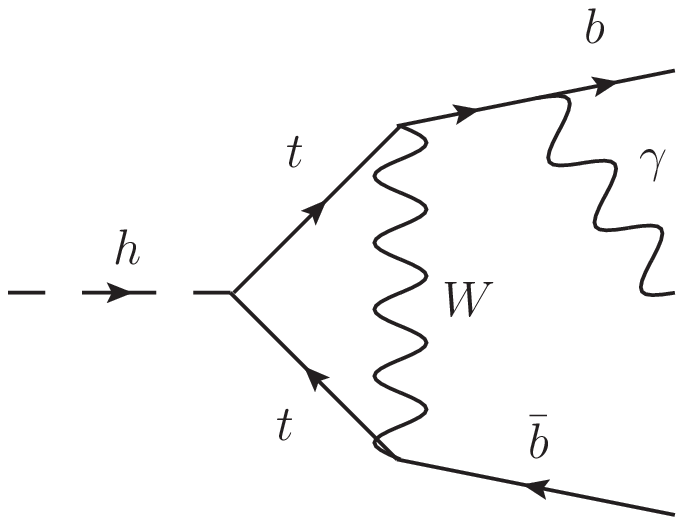}
	\caption{}
	\label{feyn:ew6}
\end{subfigure}
\begin{subfigure}[b]{0.23\textwidth}
	\includegraphics[width=\textwidth]{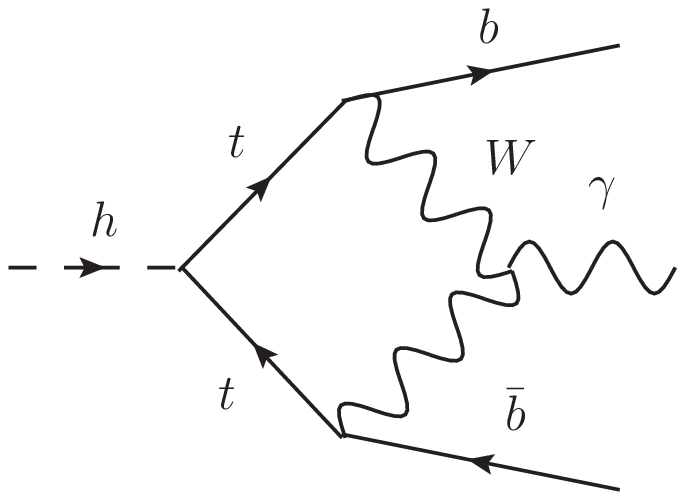}
	\caption{}
	\label{feyn:ew7}
\end{subfigure}
\begin{subfigure}[b]{0.23\textwidth}
	\includegraphics[width=\textwidth]{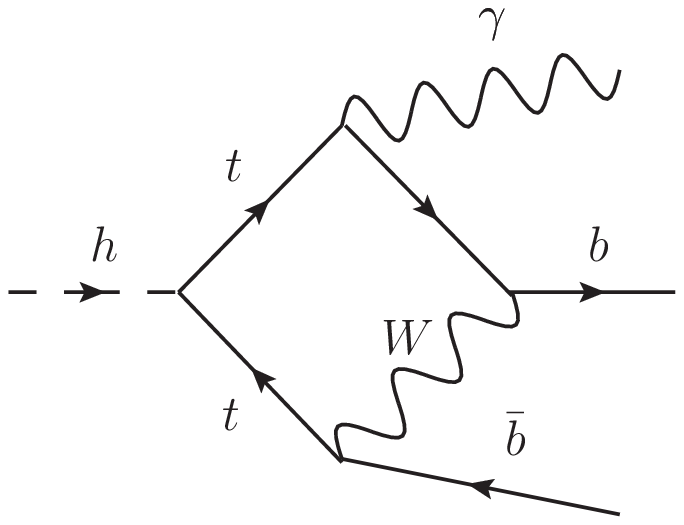}
	\caption{}
	\label{feyn:ew8}
\end{subfigure}
\caption{Representative Feynman diagrams of $h\rightarrow f\bar{f} \gamma$ with electroweak one-loop. (f)-(h) are present only in $h\rightarrow b\bar{b}\gamma$ channel.}
\label{feyn:ew}
\end{figure}

Besides the $\mathcal{O}(y_f^2\alpha)$ corrections from the chirality-flipping diagrams governed by the Yukawa couplings, the decay of a Higgs boson to a pair of fermions plus a photon can also be induced by electroweak loops of top quark and gauge bosons. Figure~\ref{feyn:ew} shows some representative electroweak one-loop diagrams. According to their distinctive kinematics and couplings, they can be cast into five classes:
\begin{enumerate}[I.]
\item
$h\rightarrow \gamma Z^*\rightarrow f\bar{f}\gamma$ (Figs.~\ref{feyn:ew1}, \ref{feyn:ew2})
\item
 $h\rightarrow \gamma \gamma^*\rightarrow f\bar{f}\gamma$  (Figs.~\ref{feyn:ew1}, \ref{feyn:ew2})
 \item
 $Z$-boson box or triangle with final state radiation (Figs.~\ref{feyn:ew3}, \ref{feyn:ew4})
 \item
 $W$-boson box or triangle with final state radiation  (Figs.~\ref{feyn:ew3}, \ref{feyn:ew4}, \ref{feyn:ew5})
 \item
top-quark box or triangle with final state radiation (Figs.~\ref{feyn:ew6}, \ref{feyn:ew7}, \ref{feyn:ew8}, only for $h\rightarrow b\bar{b}\gamma$)
\end{enumerate}
We will call them collectively the ``EW$+\gamma$'' contributions, distinctive from the chirality-flipping Yukawa corrections in Sec.~\ref{sec-QED}. The interference between the QED radiation in Fig.~\ref{feyn:qed3} and the EW$+\gamma$ processes in Fig.~\ref{feyn:ew} is suppressed by $m_f/M_W$, as they have different chiral structures for the final state fermions. The EW$+\gamma$ loops are finite at the ultra-violet (UV) so that there is no need for renormalization, as pointed out in Ref.~\cite{Abbasabadi:1995rc}. 

In the infrared (IR) limit, the amplitude in Fig.~\ref{feyn:ew} is proportional to the fermion mass $m_f$ due to the chiral structure  and the QED Ward-Takahashi identity. This is also true in the collinear region for diagrams in Figs.~\ref{feyn:ew3} and \ref{feyn:ew6}, where the amplitude factorizes into that of $h\rightarrow f\bar{f}$ convolved with a collinear splitting. Therefore, the IR/collinear singularities do not show up in the massless limit $m_f\rightarrow 0$. This behavior of Fig.~\ref{feyn:ew} remains to be valid to all orders in perturbation theory because of the chiral symmetry.
%
%
In the limit $m_f\rightarrow 0$, however, the diagrams in Figs.~\ref{feyn:ew1} and \ref{feyn:ew2} diverge as the invariant mass of the fermion pair approaches the photon pole $M_{f\bar{f}}\rightarrow0$. Therefore, a finite fermion mass needs to be kept for Figs.~\ref{feyn:ew1} and \ref{feyn:ew2} so that $M^2_{f\bar{f}} > 4m_f^2$, to regularize the divergent behavior. 

\begin{table}[tb]
\begin{center}
\renewcommand{\arraystretch}{1.5}
\begin{tabular}{|c||c|c|c|c|}
\hline
& \multicolumn{2}{c|}{Inclusive corrections}& \multicolumn{2}{c|}{Exclusive decay}\\\hline
Decay & $\delta \Gamma\ (y_f^2\alpha)$ & $\delta\Gamma\ (y_t^2\alpha^3, \alpha^4)$ & $\Gamma(f\bar{f}\gamma)$ [keV]   & BR$(f\bar{f}\gamma)$ [$10^{-4}$]\\
Channels & [keV] & [keV] &$E_\gamma^{\rm cut}=5/15$ GeV & $E_\gamma^{\rm cut}=5/15$ GeV\\
\hline\hline
$h\rightarrow b\bar{b} (\gamma)$ & $-25.3$ & 0.99  & 9.45/5.44 & 23/13 \\\hline
$h\rightarrow c\bar{c} (\gamma)$ & $-1.17$ & 0.91  & 2.48/1.73 & 6.1/4.2\\\hline\hline
$h\rightarrow \tau^+\tau^- (\gamma)$ & $-1.37$ & 0.31 & 10.4/5.63 & 25/14 \\\hline
$h\rightarrow \mu^+\mu^- (\gamma)$ & $-4.72\times 10^{-2}$ & 0.41 & 0.436/0.420 & 1.1/1.0 \\\hline
$h\rightarrow e^+e^- (\gamma)$ & $-1.29\times 10^{-6}$ & 0.60 & 0.589/0.588 & 1.4/1.4 \\\hline
\end{tabular}
\end{center}
\caption{One-loop Yukawa and EW$+\gamma$ corrections to Higgs fermionic decays. The first two columns are the inclusive corrections at the order $\mathcal{O}(y_f^2\alpha)$ and at $\mathcal{O}(y_t^2\alpha^3, \alpha^4)$, respectively. The widths and branching fractions for the exclusive decay are shown in the last two columns ($E_\gamma > 5/15~{\rm GeV}$, and $\Delta R_{f\gamma} > 0.4$). The Higgs total width of 4.1~MeV is used to calculate BRs.}
\label{tab_width}
\end{table}

We perform the calculation in the Feynman gauge. As a cross check, the analytic results have been calculated and given in \cite{Abbasabadi:1996ze}, where a non-linear $R_\xi$ gauge was used. 
All the diagrams are generated by {\it FeynArts} \cite{Hahn:2000kx}, and {\it FeynCalc} \cite{Mertig:1990an} is used to simplify the amplitudes further. The numerical evaluation of all Passarino-Veltman loop integrals \cite{Passarino:1978jh} are performed by LoopTools \cite{Hahn:1998yk}. And we use {\it Vegas} \cite{Hahn:2004fe} as the phase space integrator.

\subsection{Partial decay widths}
\label{section:decay_width}

The Yukawa corrections as in Figs.~\ref{feyn:qed1}$-$\ref{feyn:qed6} are of the order $y_f^2 \alpha$, governed by the Yukawa couplings, while the EW$+\gamma$ loops in Figs.~\ref{feyn:ew1}, \ref{feyn:ew6}$-$\ref{feyn:ew8}  involve $t\bar t h$ coupling and are thus of the order $y_t^2 \alpha^3$, and the order of $\alpha^4$ for Figs.~\ref{feyn:ew2}$-$\ref{feyn:ew5}.
We present our results for these two decay mechanisms in Table \ref{tab_width}. 
The first column shows the NLO EW corrections to the Yukawa interactions as given in Eq.~(\ref{eq:fullew}). The inclusive corrections are small and negative. The second column gives the one-loop EW$+\gamma$ contributions at the order of $y_t^2 \alpha^3$ and the order of $\alpha^4$, including their interference. The dominant EW$+\gamma$ contributions are from diagrams in Figs.~\ref{feyn:ew1} and \ref{feyn:ew2}, featured by $\gamma^*, Z \to f\bar f$. The rest of the diagrams is sub-leading and contributing about a few percent. As seen, those contributions from EW$+\gamma$ loops are essentially independent of the light fermion masses and thus independent of the Yukawa couplings. The moderate dependence on the mass is due to the kinematical enhancement from the photon splitting near $M_{f\bar{f}} \sim 2m_f$. In comparison with these two decay mechanisms of the Yukawa corrections and EW$+\gamma$ contributions, we see that the orders of magnitudes are comparable for the $c\bar c$ case. The Yukawa corrections dominate over the EW$+\gamma$ contributions for the decays to $b\bar b$ and $\tau^+ \tau^-$, while it becomes the other way around for $\mu^+ \mu^- $ and $e^+ e^- $, due to their much smaller Yukawa couplings. 

\begin{figure}[tb]
\centering
\hspace{-1.cm}
\includegraphics[width=0.7\textwidth]{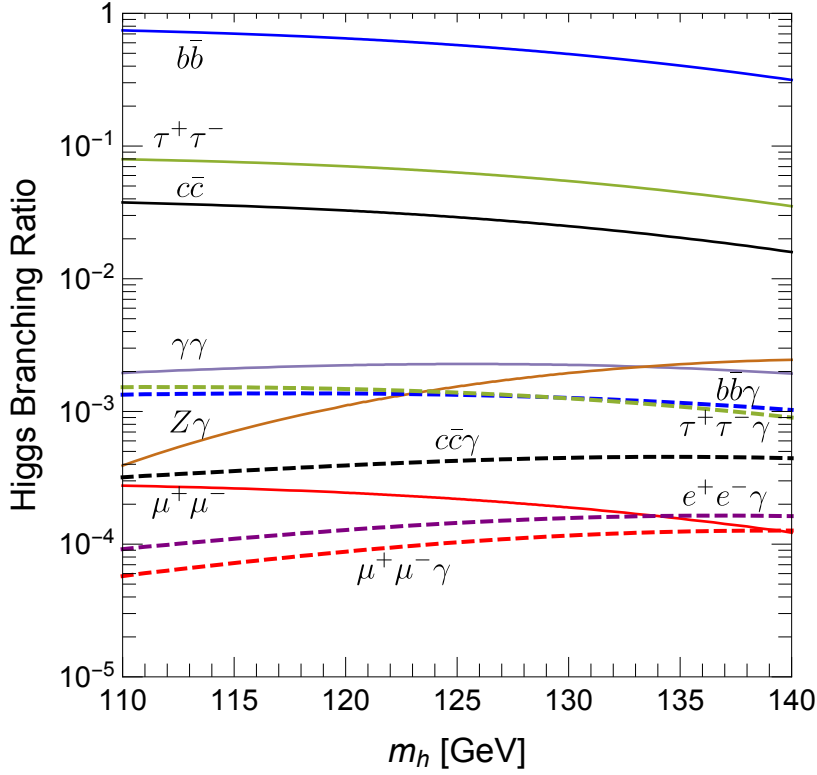}
\caption{SM Higgs decay branching fractions to fermions with and without the additional photon $E_\gamma > 15~{\rm GeV}$ and $\Delta R>0.4$.}
\label{fig:br}
\end{figure}

From the observational point of view with the $f\bar f \gamma$ events, we require a photon in the final state to satisfy the minimal acceptance cuts\footnote{The kinematical variables here are in the Higgs boson rest-frame. In realistic simulations, one may need to evaluate them in the lab frame.}
\begin{equation}
E_\gamma > 5~{\rm or\ 15\ GeV}\ \ {\rm and}\ \  \Delta R_{\gamma f},\ \Delta R_{ \gamma \bar f} > 0.4 ,
\label{eq:cuts}
\end{equation}
with the separation defined in the pseudo rapidity-azimuthal angle space 
$\Delta R_{\gamma f} =(\Delta\eta^2 + \Delta\phi^2)^{1/2}$.
In Table \ref{tab_width}, we list the partial widths and the branching fractions (BR) in the last two columns with a photon satisfying the cuts in Eq.~(\ref{eq:cuts}). 
We note that the exclusive partial widths of $f\bar f \gamma$ can be sizable. 
The branching fractions of $b\bar b\gamma,\ \tau^+\tau^-\gamma$ are of the order of $0.2\%$, largely from the QED radiation and thus quite sensitive to the photon energy threshold. 
The branching fraction of $c\bar c\gamma$, on the other hand, is about $6\times10^{-4}$, with  comparable contributions from the QED radiation and EW$+\gamma$ processes, and thus also rather sensitive to the photon energy cut. Those for $e^+e^-\gamma$ and $\mu^+\mu^-\gamma$ are about $10^{-4}$, dominantly from the EW$+\gamma$ processes and thus insensitive to the photon energy threshold, to be further discussed below. It is interesting to note that it would be totally conceivable for observation of those clean leptonic channels at the HL-LHC. We also show the Higgs decay branching fractions to fermions in Fig.~\ref{fig:br}. It is quite informative to compare our results for the exclusive radiative decays $h\to f\bar f \gamma$ with those from $h\to f\bar f$.

\begin{figure}[tb]
\centering
\begin{subfigure}{0.32\textwidth}
\includegraphics[width=\textwidth]{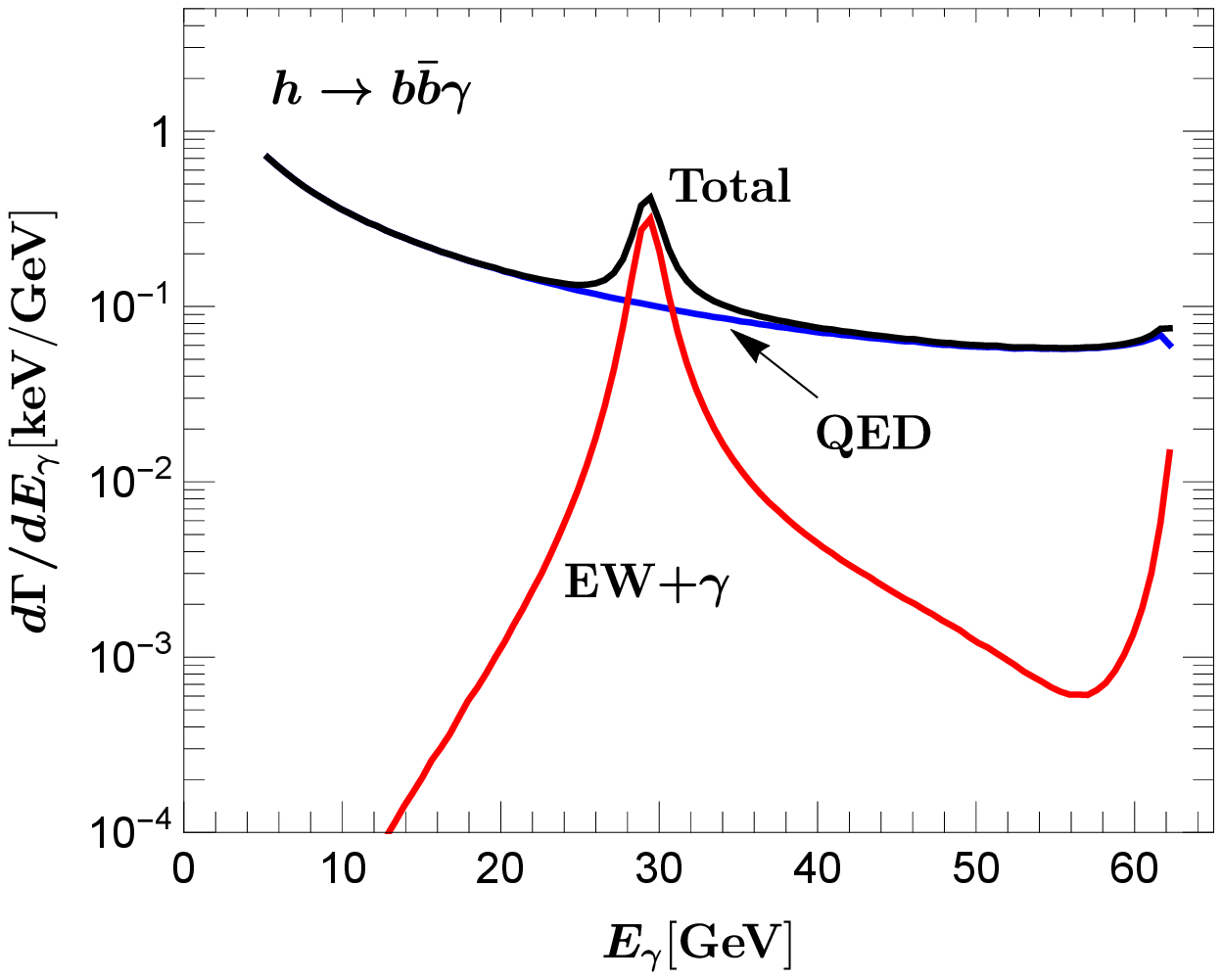}
\caption{}
\label{plot:decay_Ea_b}
\end{subfigure}
\begin{subfigure}{0.32\textwidth}
\includegraphics[width=\textwidth]{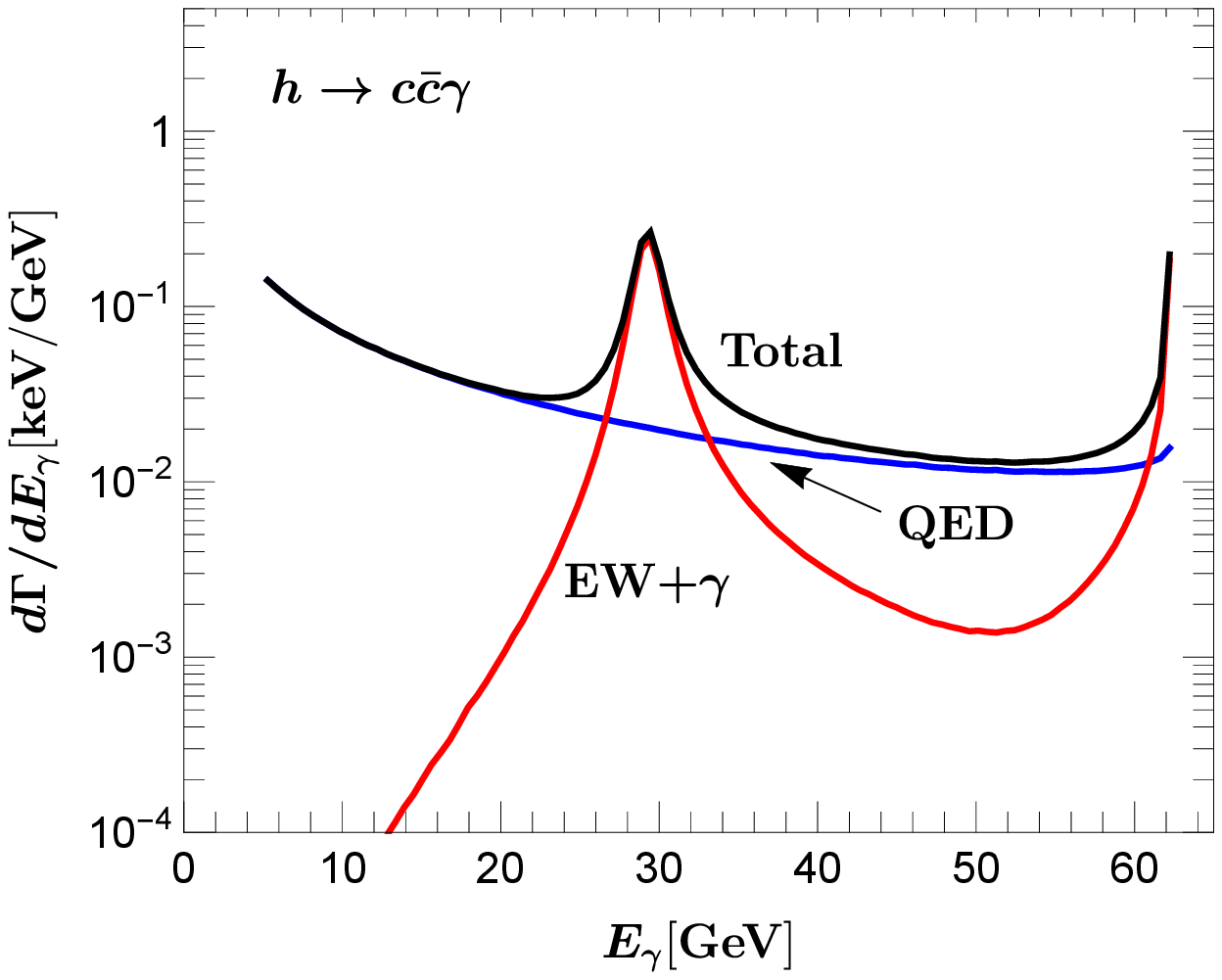}
\caption{}
\label{plot:decay_Ea_c}
\end{subfigure}
\\
\begin{subfigure}{0.32\textwidth}
\includegraphics[width=\textwidth]{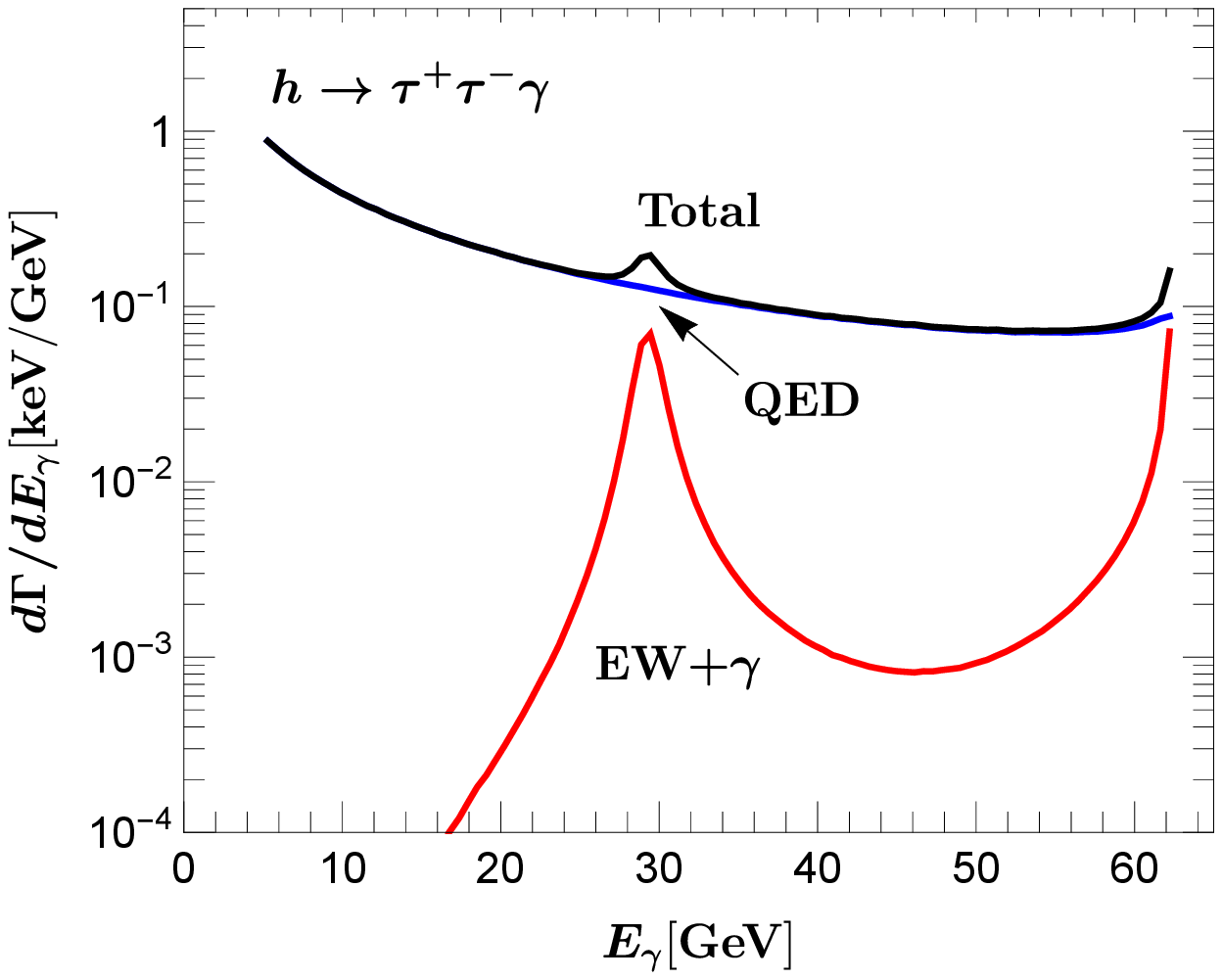}
\caption{}
\label{plot:decay_Ea_t}
\end{subfigure}
\begin{subfigure}{0.32\textwidth}
\includegraphics[width=\textwidth]{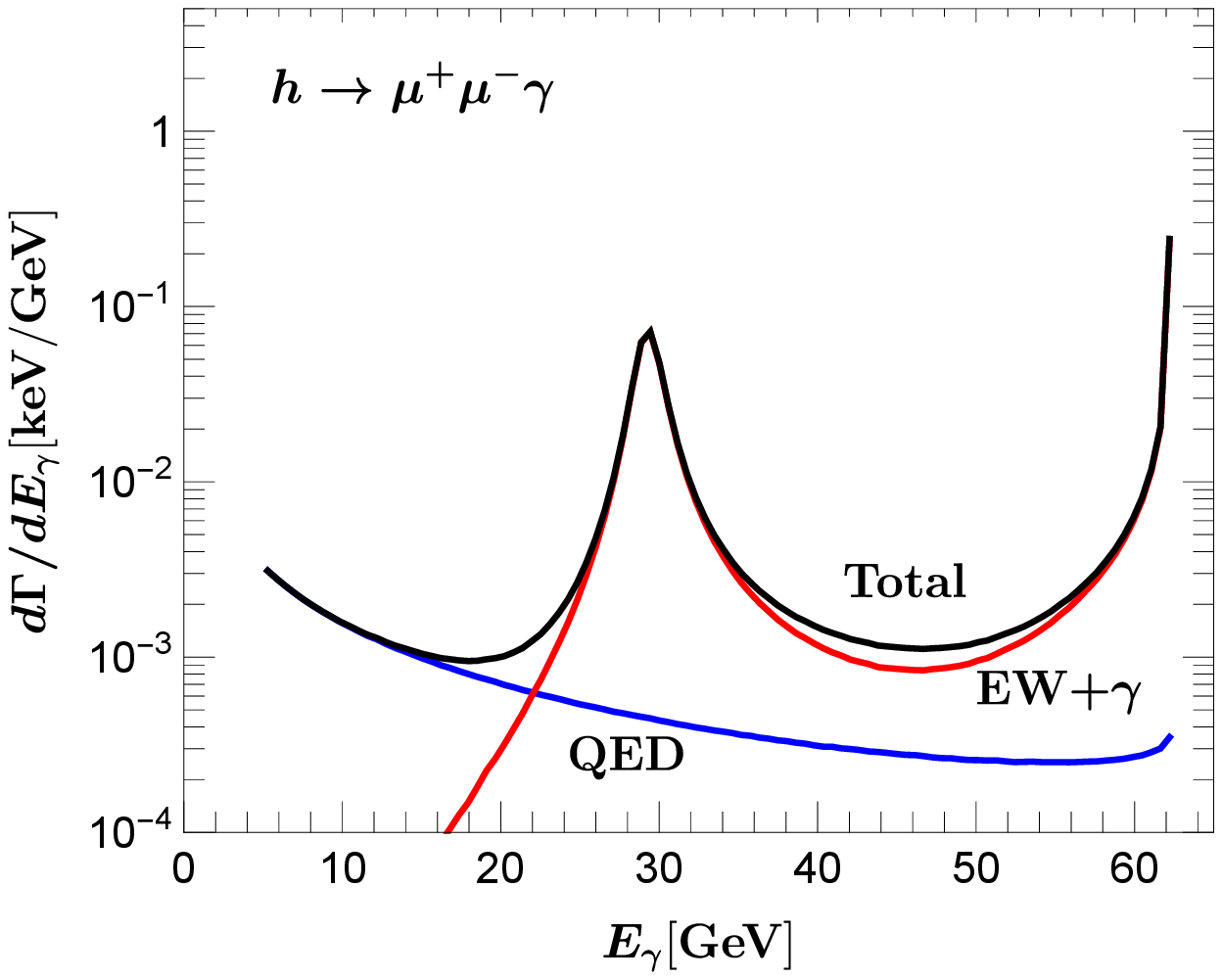}
\caption{}
\label{plot:decay_Ea_m}
\end{subfigure}
\begin{subfigure}{0.32\textwidth}
\includegraphics[width=\textwidth]{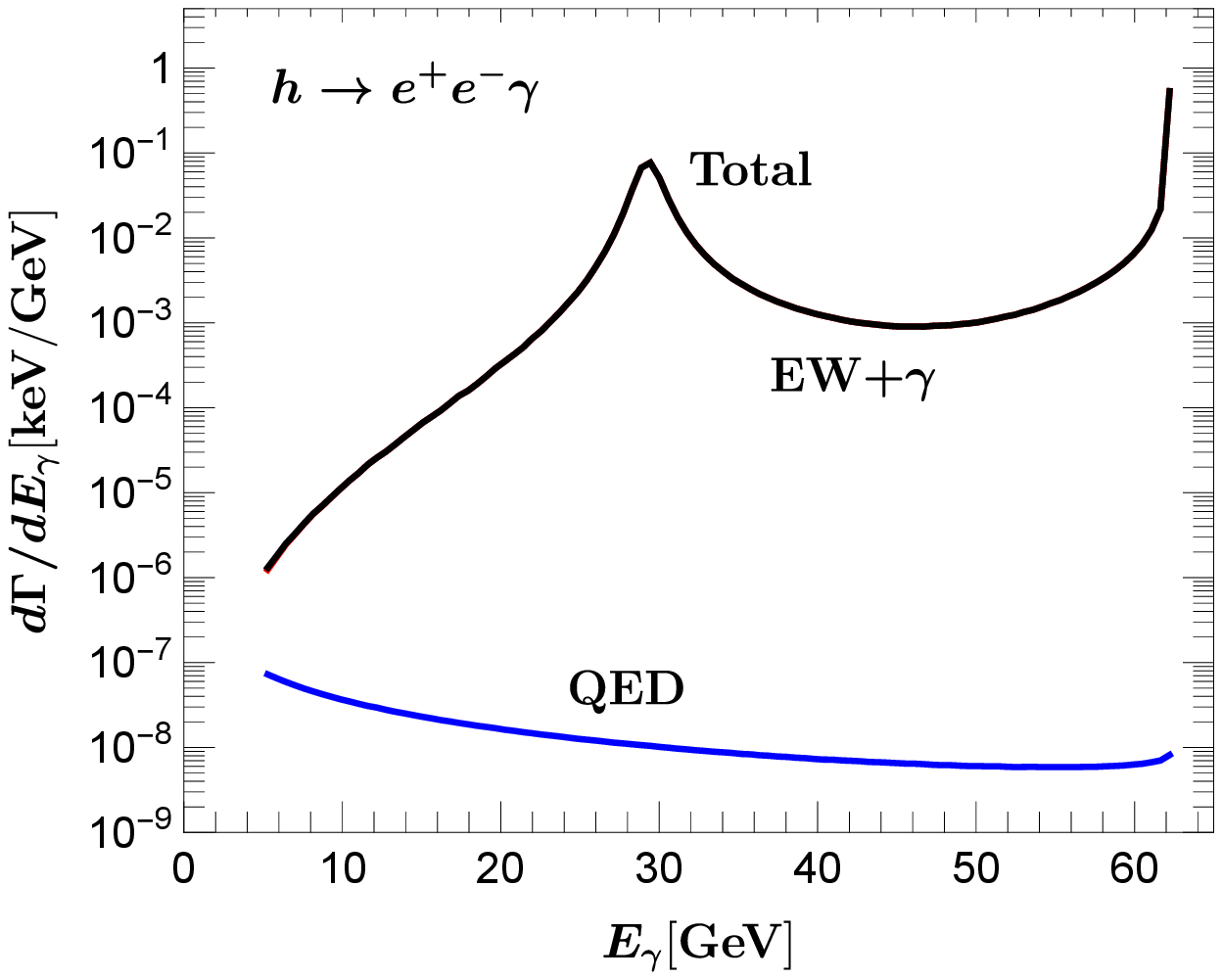}
\caption{}
\label{plot:decay_Ea_e}
\end{subfigure}
\caption{The photon energy distributions in $h\rightarrow f\bar{f}\gamma$ $(f = b, c, \tau, \mu, e)$ in the Higgs boson rest frame. The blue curves are for the QED radiation (Fig.~\ref{feyn:qed3}); the red curves are for the EW$+\gamma$ processes (Fig.~\ref{feyn:ew}); the upper black lines are for the total.}
\label{plot:decay_Ea}
\end{figure}

It is interesting to explore some kinematical distributions to appreciate the underlying decay mechanisms and to guide future experimental searches. In Fig.~\ref{plot:decay_Ea}, we show the photon energy distributions in the Higgs boson rest frame for the individual fermionic channels for the QED radiation (solid blue curves) and for the EW$+\gamma$ processes (solid red curves) and the total (upper curves). The $E_\gamma$ spectrum of the QED radiation exhibits the common infrared behavior: the observable photon energy spectrum diverges like $dE_\gamma/E_\gamma$, although the inclusive integrated rate is finite due to the cancelation from the virtual loop diagrams. 
The energy spectrum of the EW$+\gamma$ processes, on the other hand, exhibits a double-hump structure as seen from the red curves in Fig.~\ref{plot:decay_Ea}, characterizing the two dominant underlying processes
\begin{eqnarray}
\label{eq:ez}
E_\gamma&=& {m_h \over 2}(1-{m_Z^2\over m_h^2}) \approx 30~{\rm GeV,\ for}\ \ \gamma Z\ \ {\rm production},\\
E_\gamma &=& {m_h \over 2}(1-{m_{\gamma^*}^2\over m_h^2}) \approx 63~{\rm GeV, \ for}\ \ \gamma \gamma^*\ \ {\rm production}.
\label{eq:eg}
\end{eqnarray}
The diagrams of Figs.~\ref{feyn:ew3}~and~\ref{feyn:ew5} have a spurious divergence in the infrared (soft) and collinear region. However, in the soft/collinear limit, the amplitude has to be proportional to the fermion mass due to conservation of angular momentum, and thus vanishes in the massless limit, as confirmed by the plots here.

\begin{figure}[tb]
\centering
\begin{subfigure}{0.32\textwidth}
\includegraphics[width=\textwidth]{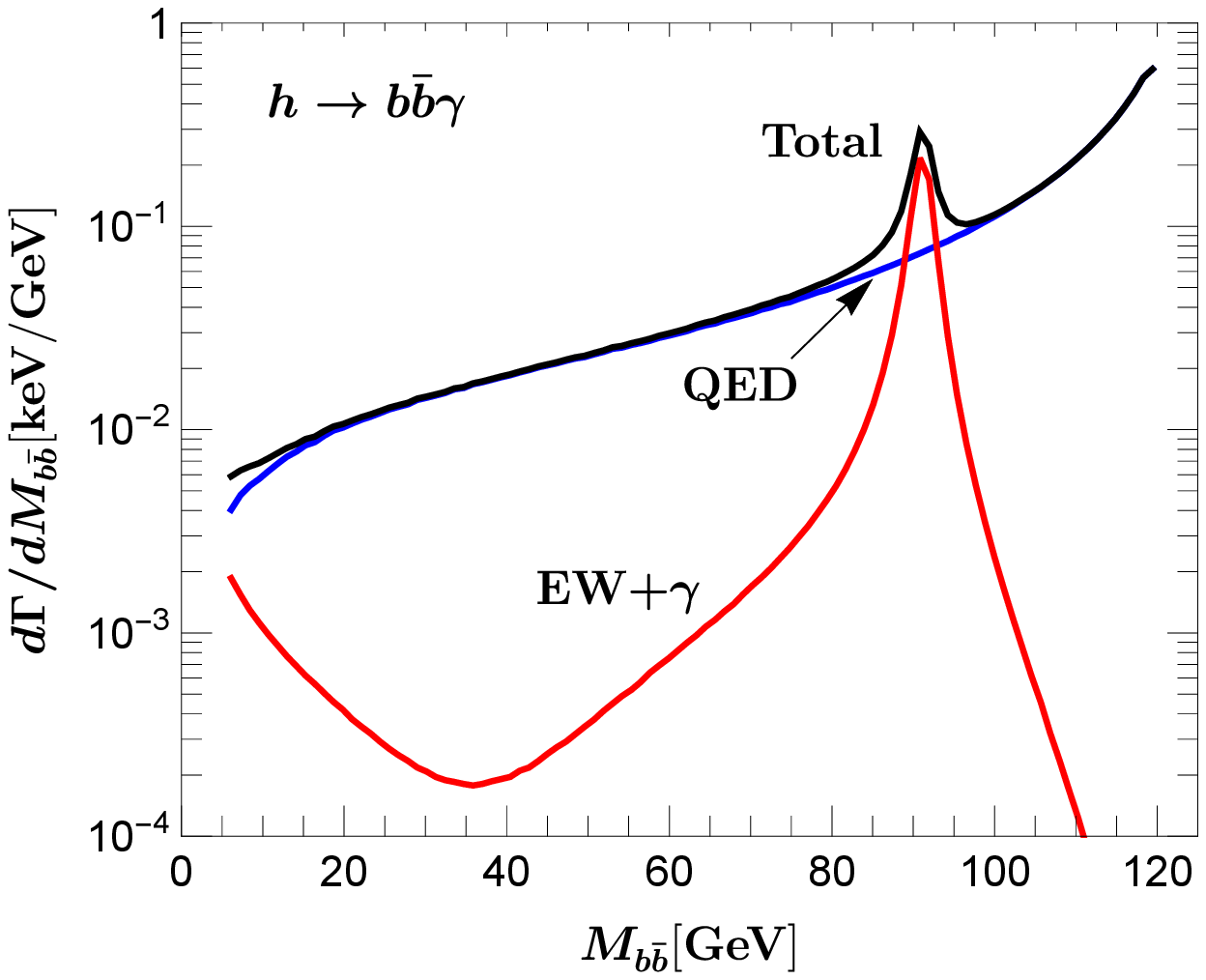}
\caption{}
\label{plot:decay_Mbb}
\end{subfigure}
\begin{subfigure}{0.32\textwidth}
\includegraphics[width=\textwidth]{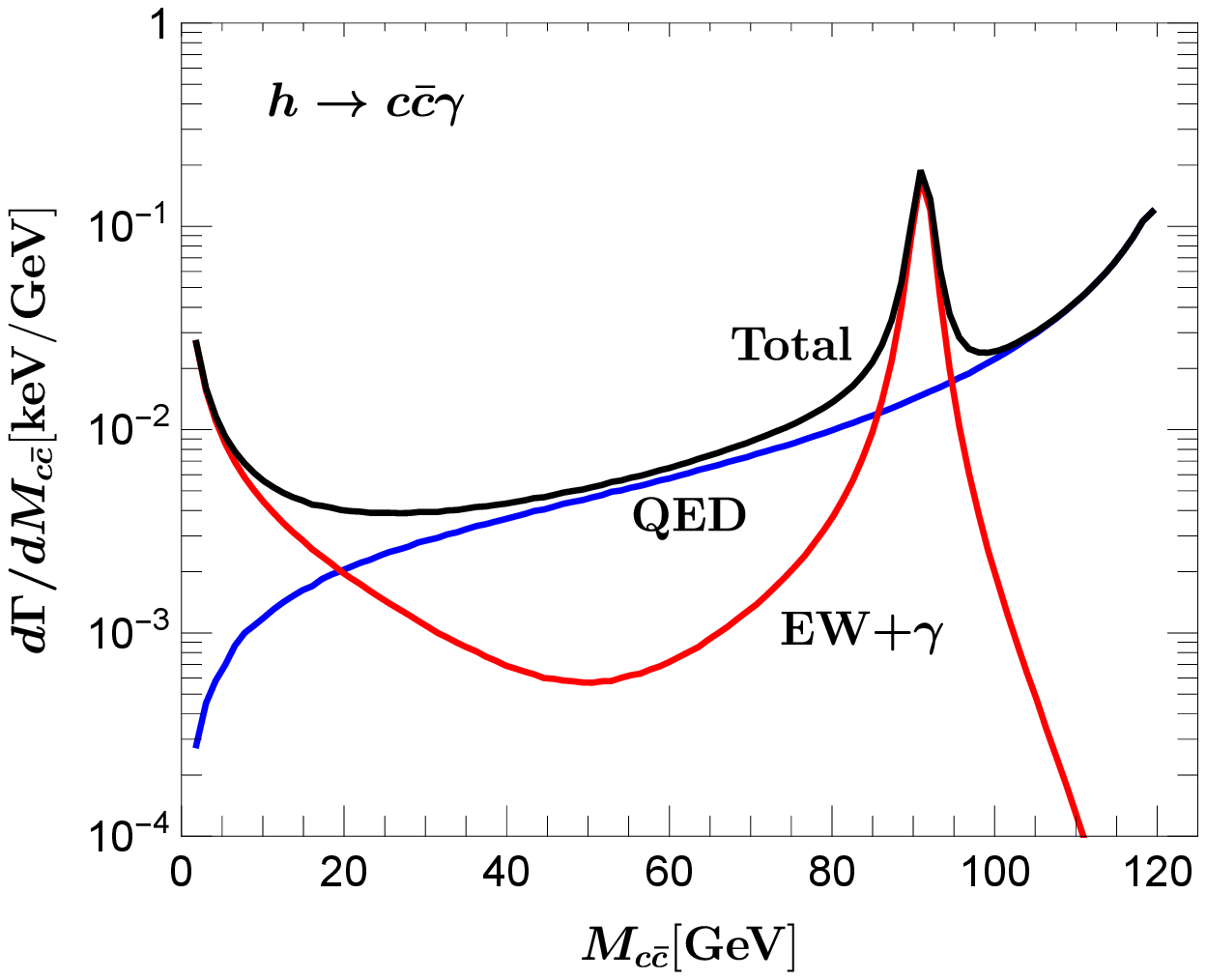}
\caption{}
\label{plot:decay_Mcc}
\end{subfigure}
\\
\begin{subfigure}{0.32\textwidth}
\includegraphics[width=\textwidth]{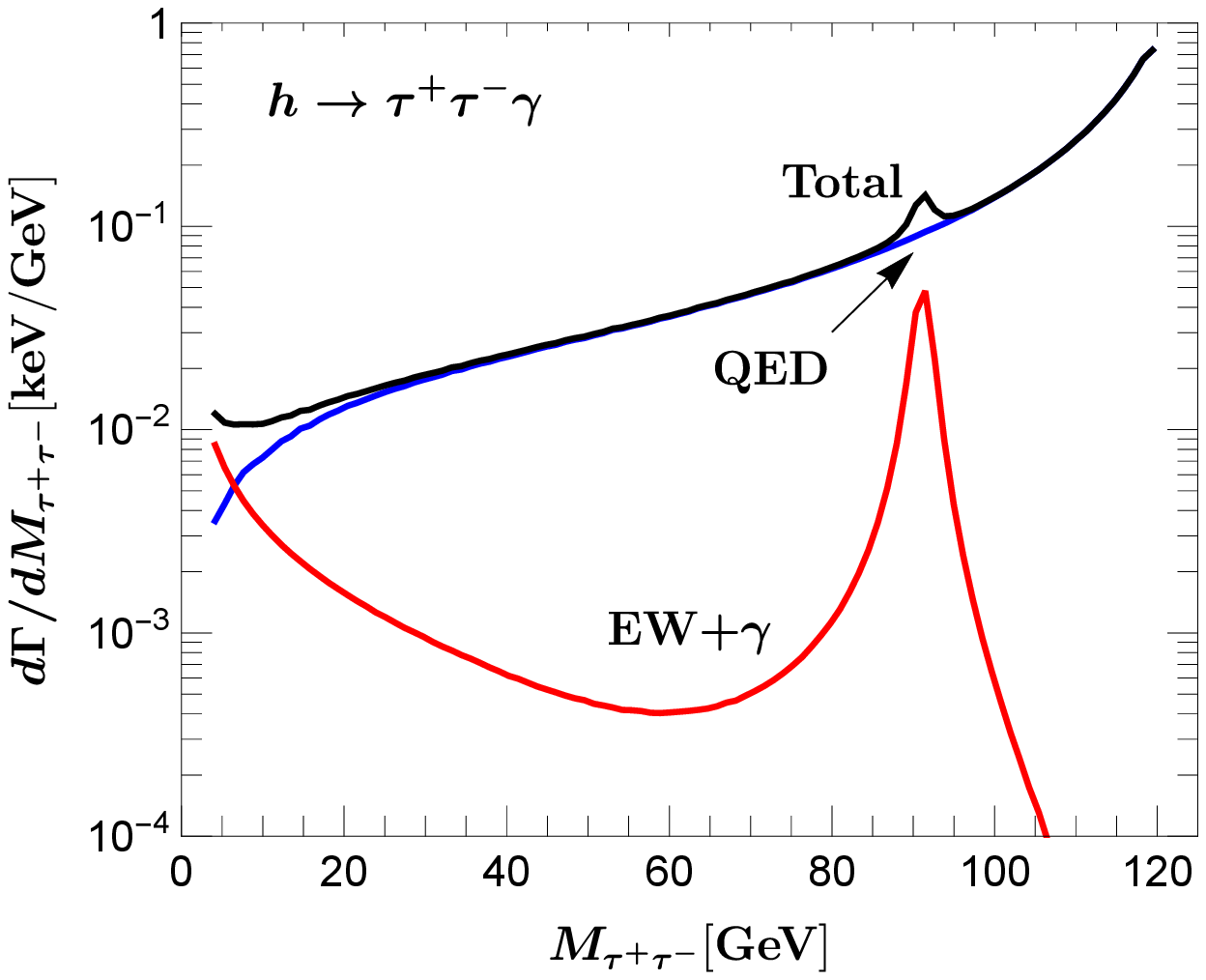}
\caption{}
\label{plot:decay_Mtt}
\end{subfigure}
\begin{subfigure}{0.32\textwidth}
\includegraphics[width=\textwidth]{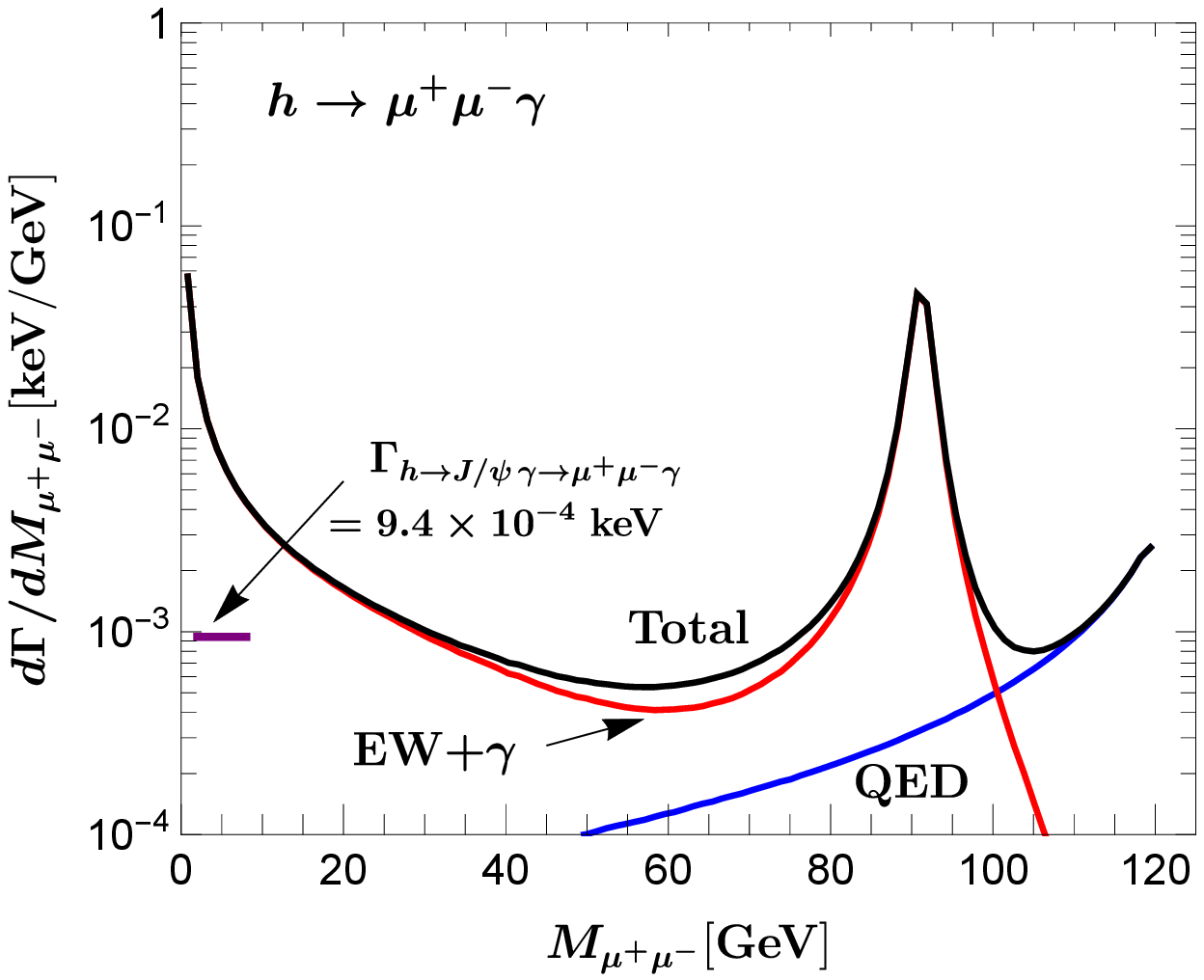}
\caption{}
\label{plot:decay_Mmm}
\end{subfigure}
\begin{subfigure}{0.32\textwidth}
\includegraphics[width=\textwidth]{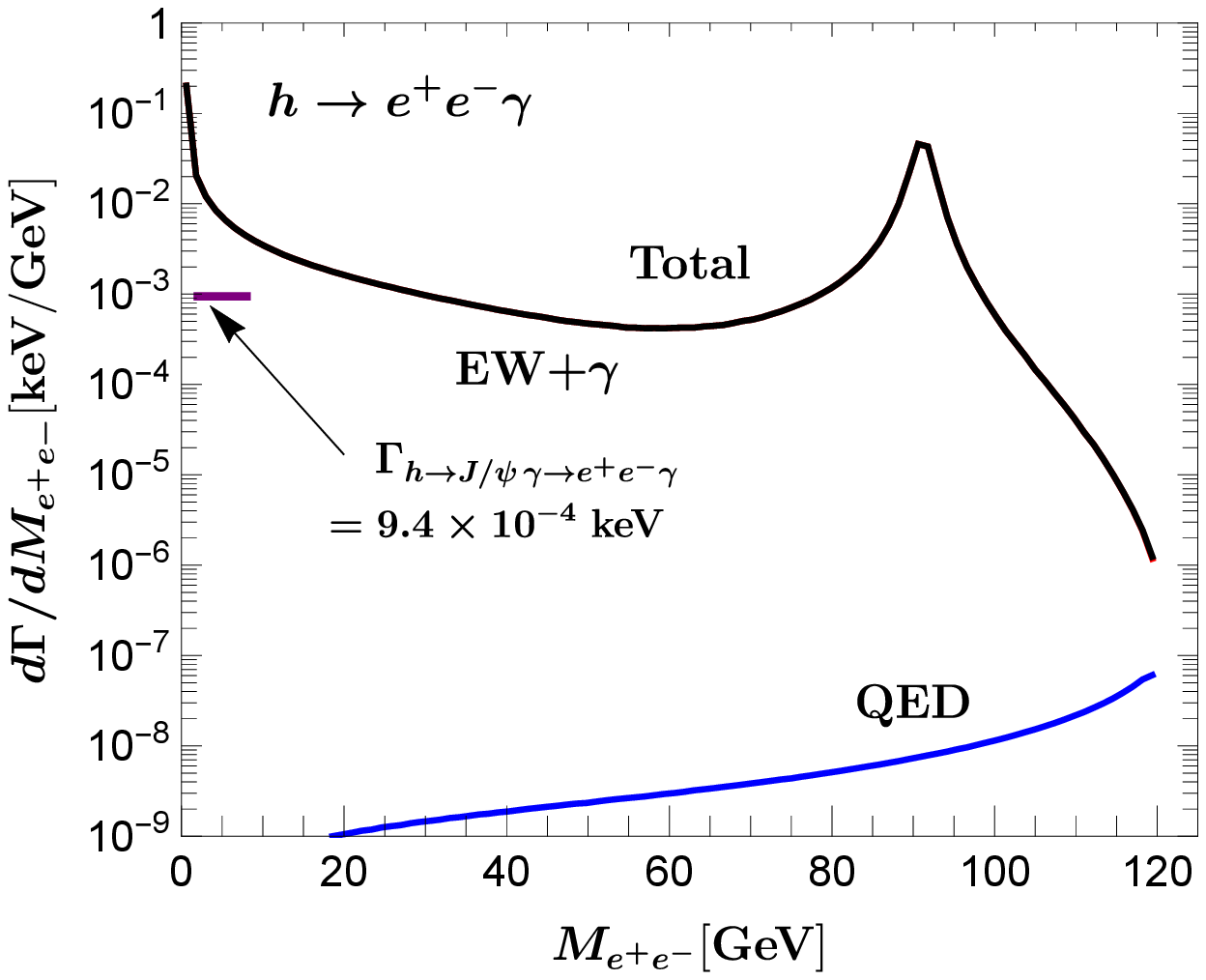}
\caption{}
\label{plot:decay_Mee}
\end{subfigure}
\caption{The invariant mass distributions of the fermion pair in $h\rightarrow f\bar{f}\gamma$ $(f = b, c, \tau, \mu, e)$. The blue curves are for the QED radiation (Fig.~\ref{feyn:qed3}); the red curves are for the EW$+\gamma$ processes (Fig.~\ref{feyn:ew}); the upper black lines are for the total. 
The decay widths for the channels $h\to J\psi\ \gamma \to \ell^+\ell^- \gamma$ are indicated by the horizontal bars in (d) and (e), in units of keV without the photon acceptance cuts.}
\label{plot:decay_Mff}
\end{figure}

We also show the invariant mass distributions of the fermion pairs in Fig.~\ref{plot:decay_Mff}. Generally speaking, there is a correlation between the invariant mass and the energy as $M_{ff}^2 = m_h^2 - 2m_h E_\gamma$. While the invariant mass spectrum of the QED radiation has a rather smooth distribution, those from EW$+\gamma$ processes are again seen with the double-humps, one near the $Z$-pole and another near $m_{\gamma^*} \sim 2m_f$, which becomes more pronounced for a smaller fermion mass. This is the reason why the decay rate for $e^+e^-\gamma$ is larger than that for $\mu^+\mu^-\gamma$.

\begin{figure}[tb]
	\centering
	\begin{subfigure}{0.32\textwidth}
		\includegraphics[width=\textwidth]{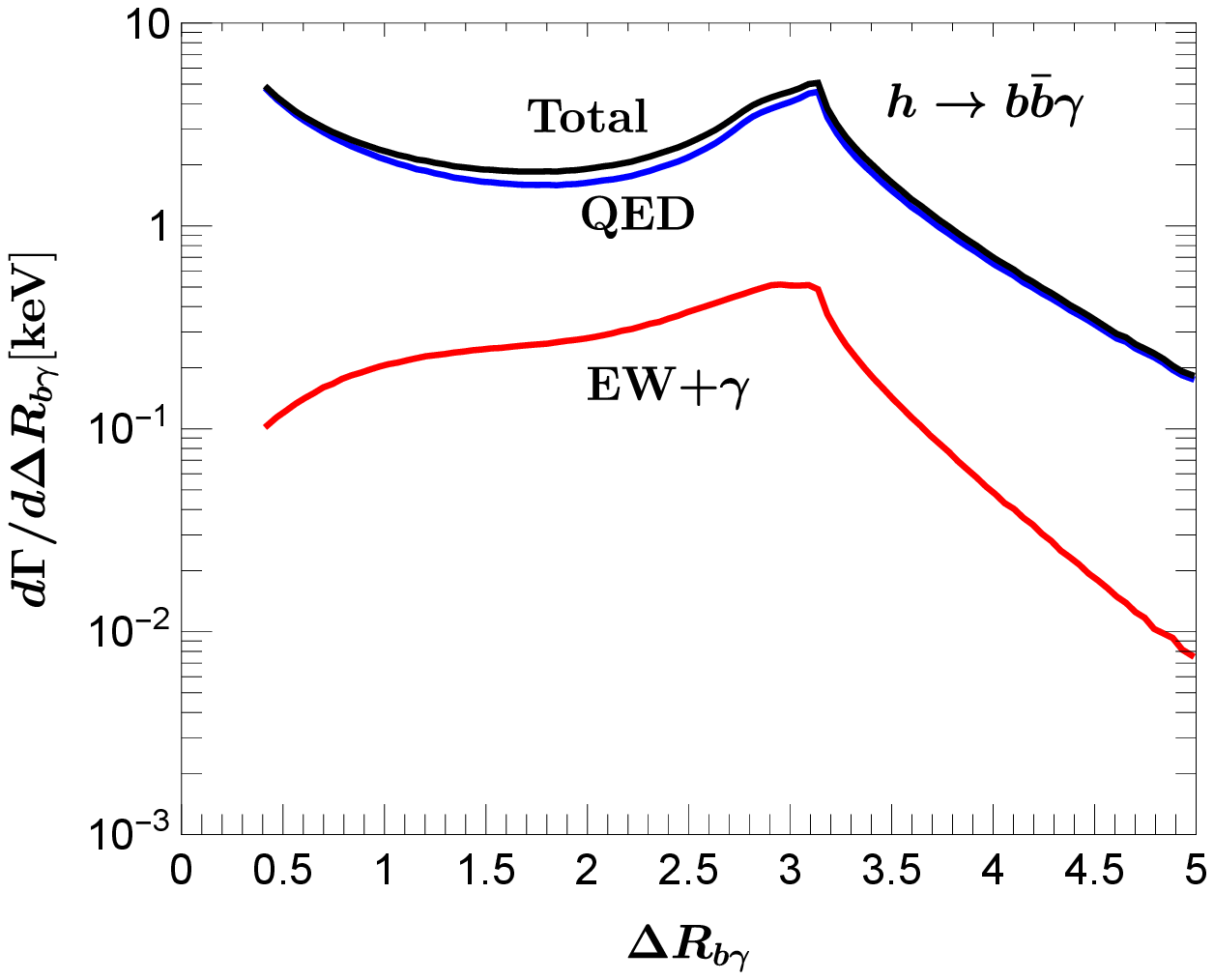}
		\caption{}
		\label{plot:decay_dRba}
	\end{subfigure}
	\begin{subfigure}{0.32\textwidth}
		\includegraphics[width=\textwidth]{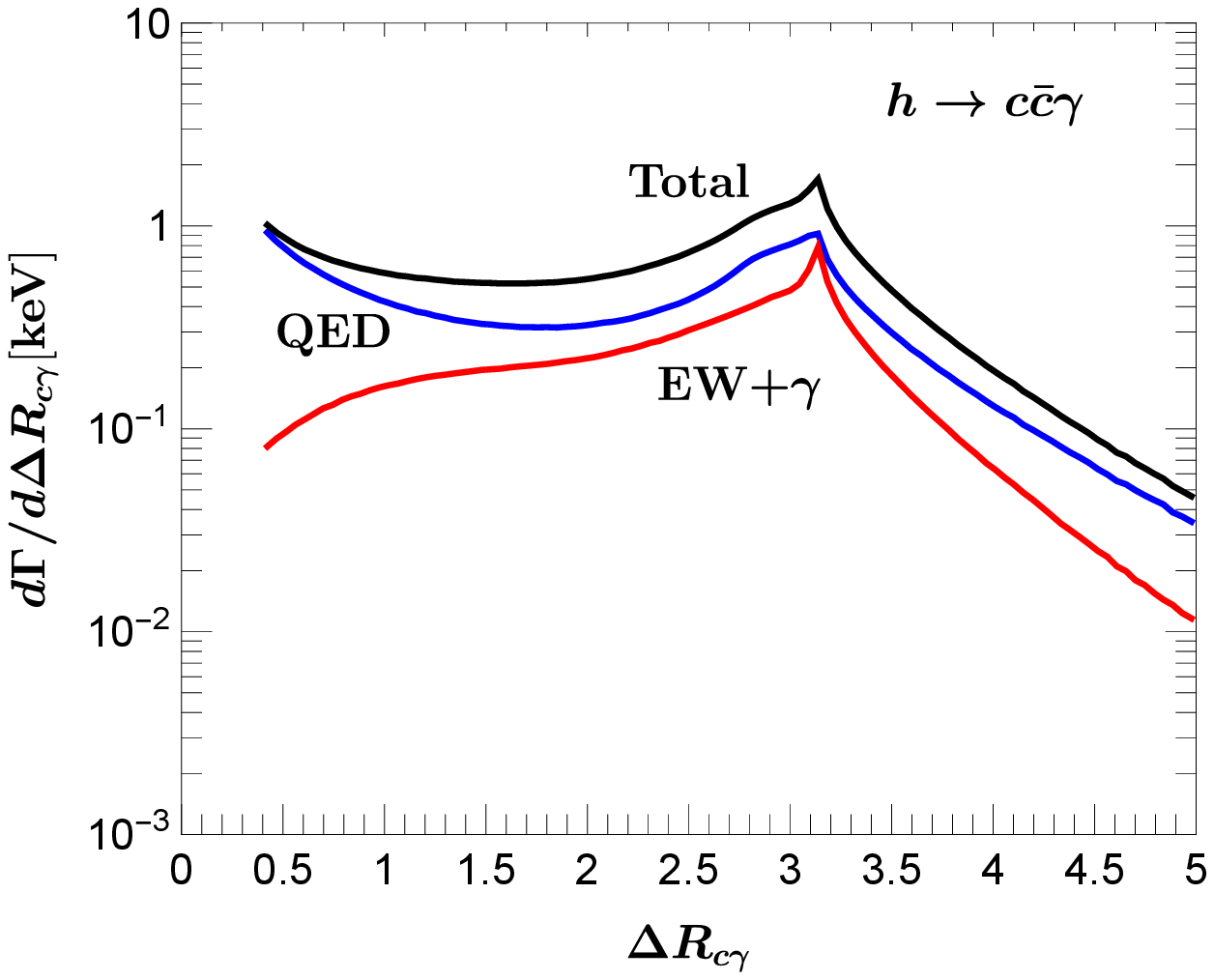}
		\caption{}
		\label{plot:decay_dRca}
	\end{subfigure}
\\
	\begin{subfigure}{0.32\textwidth}
		\includegraphics[width=\textwidth]{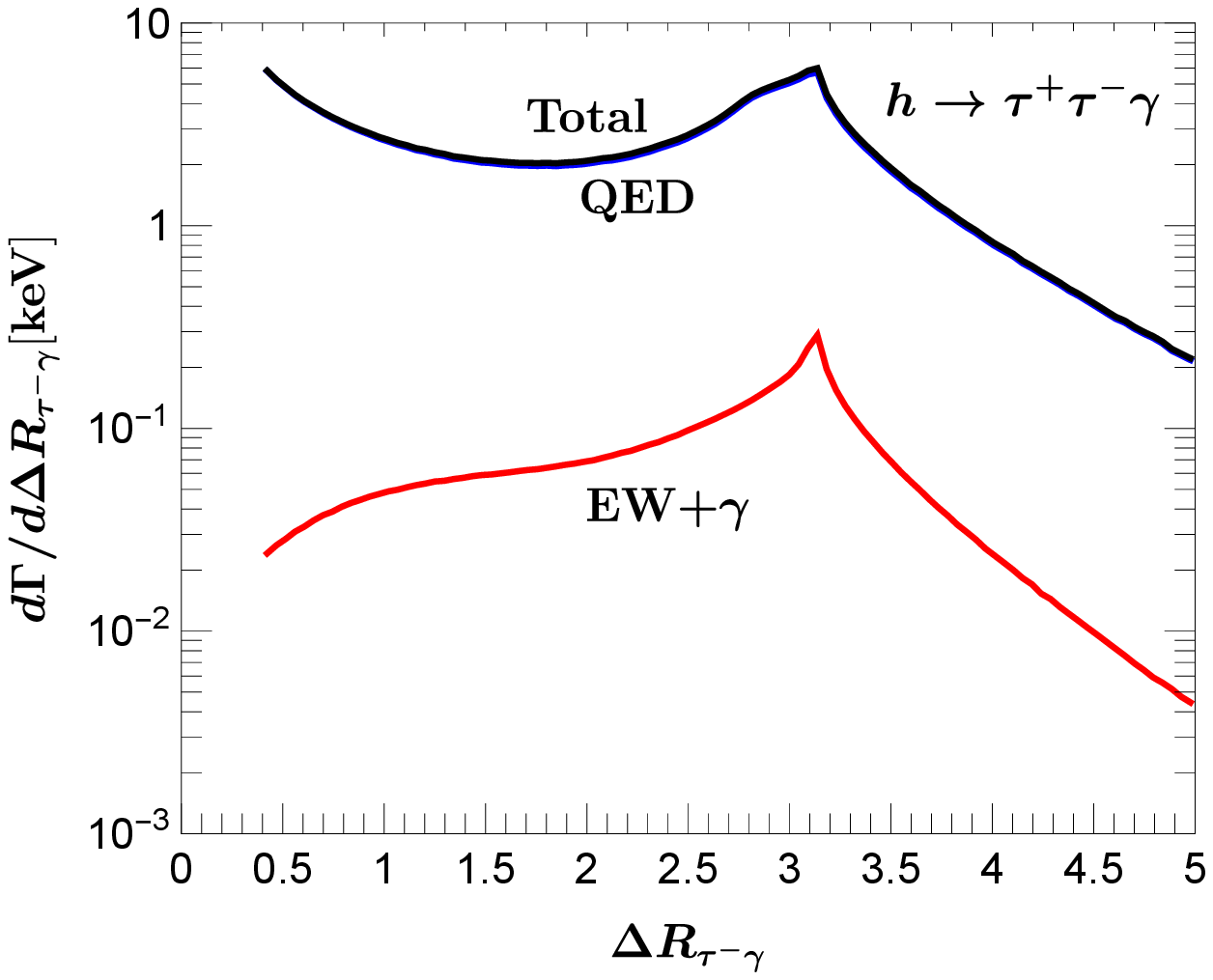}
		\caption{}
		\label{plot:decay_dRta}
	\end{subfigure}
	\begin{subfigure}{0.32\textwidth}
		\includegraphics[width=\textwidth]{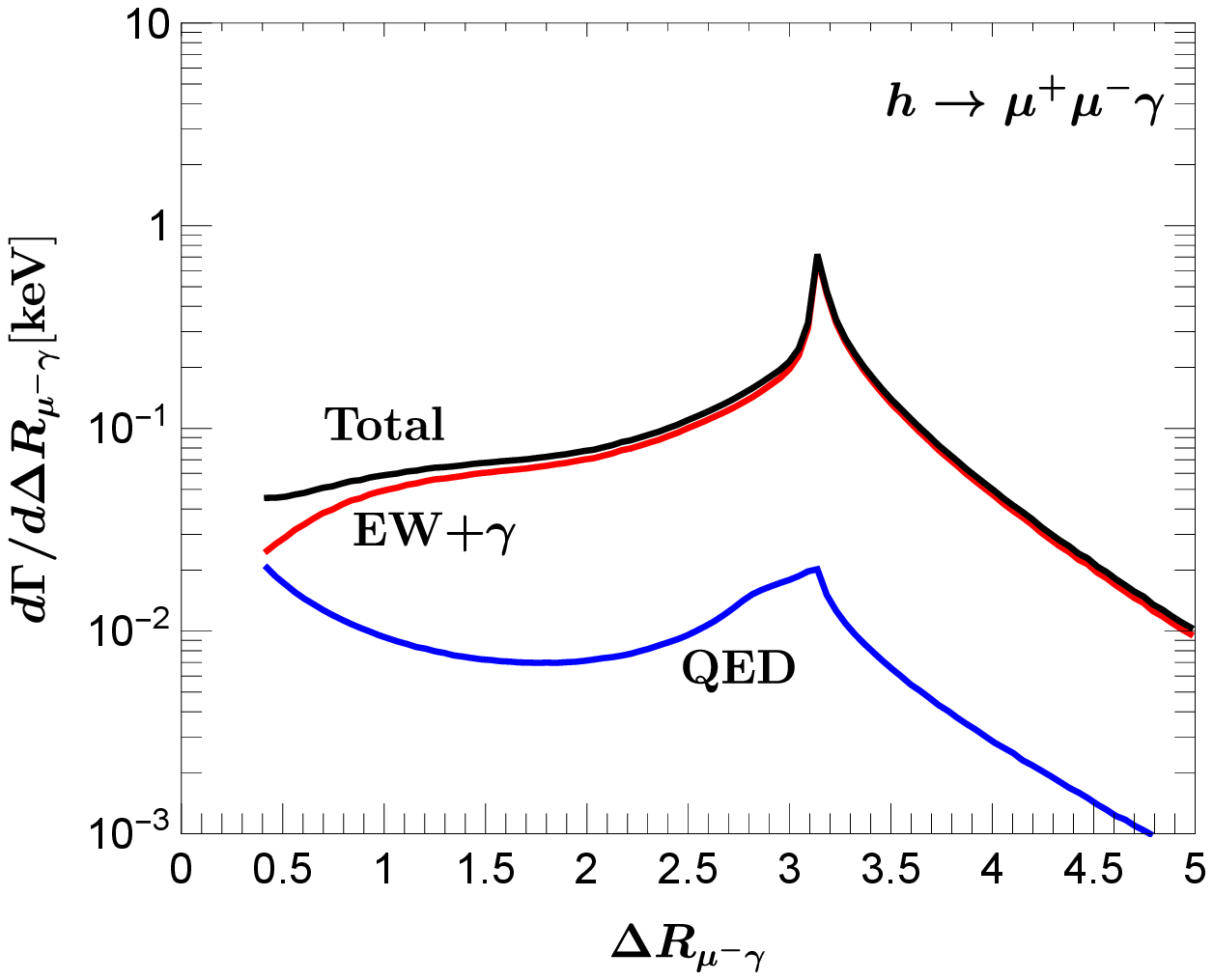}
		\caption{}
		\label{plot:decay_dRma}
	\end{subfigure}
	\begin{subfigure}{0.32\textwidth}
		\includegraphics[width=\textwidth]{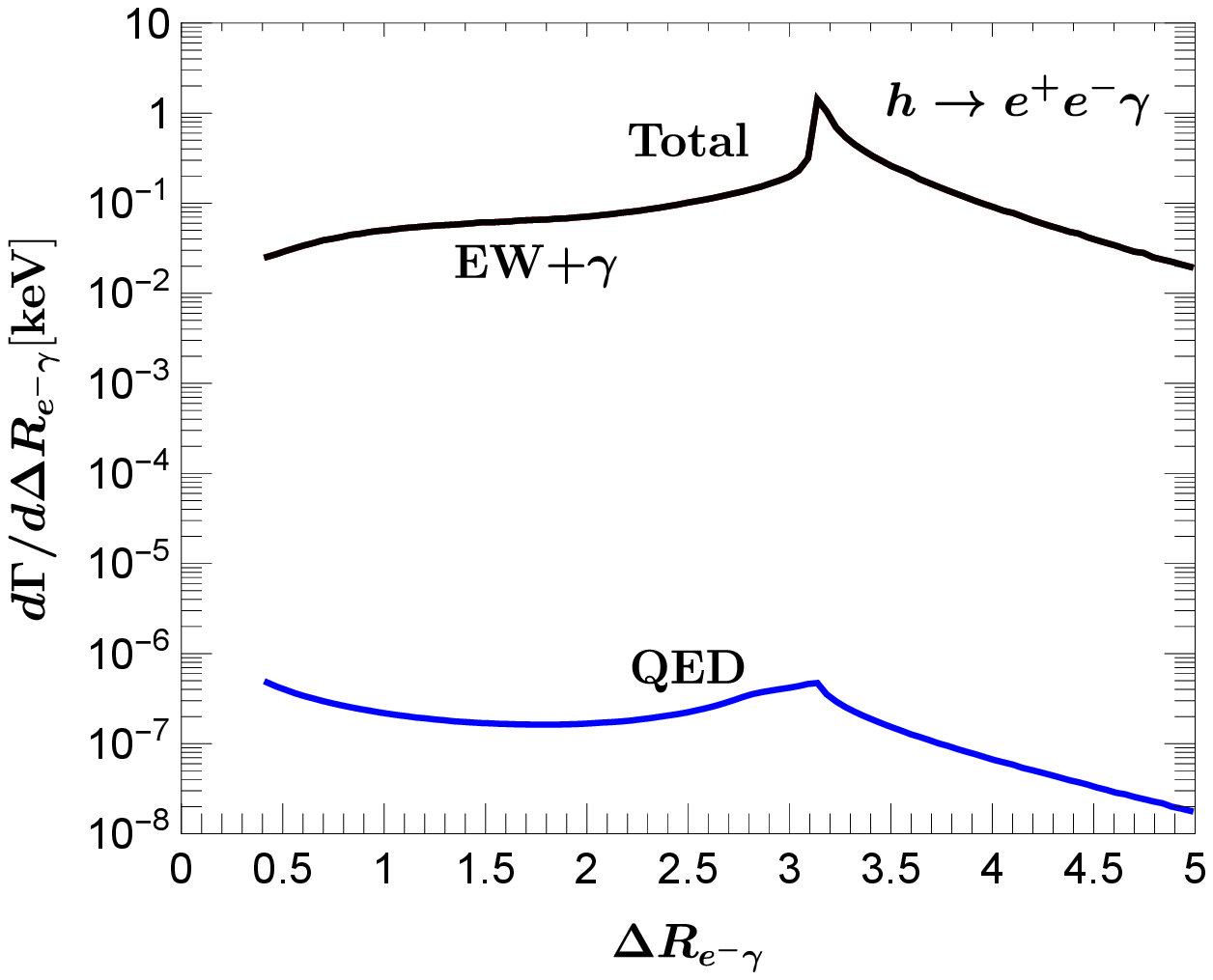}
		\caption{}
		\label{plot:decay_dRea}
	\end{subfigure}
	\caption{The distributions of the photon separation from the fermions in $h\rightarrow f\bar{f}\gamma$ $(f = b, c, \tau, \mu, e)$ in the Higgs boson rest frame. The blue curves are for the QED radiation (Fig.~\ref{feyn:qed3}); the red curves are for the EW$+\gamma$ processes (Fig.~\ref{feyn:ew}); the upper black lines are for the total.}
	\label{plot:decay_dRfa}
\end{figure}

Finally, we show in Fig.~\ref{plot:decay_dRfa} the distributions of the photon separation from the fermions, defined in Eq.~(\ref{eq:cuts}).
As expected, the QED radiation exhibit a collinear divergence near $\Delta R_{\gamma f} \to 0$, and the EW$+\gamma$ processes lead to a back-to-back structure $\Delta R_{\gamma f} \to \pi$.

\section{LHC Search for $\ell^+ \ell^- \gamma$}
\label{sec:LHCll}

In the upcoming and future LHC programs, it is of fundamental importance to observe the Higgs boson rare decays to check the consistency of the SM and seek for hints for new physics. 
Given the anticipated large yield at the HL-LHC, reaching about 150 million Higgs bosons, the very clean final states $\ell^+ \ell^- \gamma\ (\ell=\mu, e)$ should be among the first to look for. We now discuss their observability at the LHC.

As mentioned in Sec.~\ref{section:decay_width}, the radiative decays $h\rightarrow \mu^+ \mu^- \gamma$ and $h\rightarrow e^+e^-\gamma$ are mainly from the chirality-conserving EW$+\gamma$ loop diagrams. As seen from Figs.~\ref{plot:decay_Mmm} and \ref{plot:decay_Mee}, the leading contributions are from $h\rightarrow  \gamma^* \gamma, Z\gamma \rightarrow \ell^+\ell^-\gamma$ \cite{Abbasabadi:2000pb, Gainer:2011aa, Chen:2012ju, Passarino:2013nka, Sun:2013rqa, Korchin:2014kha}. 
It is thus a good search strategy to focus on the $\gamma$-pole and the $Z$-pole. 
Some searches have been carried out by ATLAS \cite{Aad:2014fia} and CMS \cite{Chatrchyan:2013vaa, Khachatryan:2015lga} at the $7-8$ TeV LHC. 
We present our analyses below in the hope to serve as a theoretical guidance for the future experimental searches at the LHC. 
We focus on the leading production for the Higgs boson via the gluon fusion. 
The QCD corrections are taken into account by multiplying a flat NNLO QCD $K$-factor of $K=2.7$ for the gluon fusion \cite{Anastasiou:2016cez}.
The dominant SM background is the Drell-Yan production of the lepton pair $\ell^+\ell^-$ with an initial/final state photon radiation. We calculate the background processes at LO using {\it MadGraph} \cite{Alwall:2014hca}, and then multiplied by flat QCD $K$-factors 
 $K=1.4$ for $pp\rightarrow Z\gamma\rightarrow \ell^+\ell^-\gamma$ \cite{Grazzini:2013bna}, and $K=6.2$ for $pp\rightarrow \gamma^*\gamma\rightarrow \ell^+\ell^-\gamma$ \cite{Campbell:2016yrh}.


\subsection{$h\rightarrow \gamma^*\gamma\rightarrow\ell^+\ell^-\gamma$}
To make the close connection with the LHC searches, we first follow the event selection cuts adopted by the CMS collaboration \cite{Khachatryan:2015lga}. As the invariant mass of the lepton pair approaches to $2m_f$, the lepton pair tends to be collimated. This becomes particularly challenging for the electron channel, because the electron pair merges into one supercluster. Therefore, a single muon plus a photon trigger for the muon channel and a di-photon trigger for the electron channel are implemented. To select the signal events near the $\gamma$-pole from the Higgs decay and effectively suppress the backgrounds, we require the invariant masses to be
\begin{equation}
M_{\mu\mu} < 20~{\rm GeV}, \quad M_{ee} < 1.5~{\rm GeV}, \quad 120~{\rm GeV}<M_{\ell\ell\gamma}<130~{\rm GeV}.
\label{eq:mcut}
\end{equation}
The leading (sub-leading) muon must satisfy the acceptance of the transverse momentum and pseudo-rapidity
\begin{equation}
p_T^\mu > 23~(4)~{\rm GeV},\quad |\eta_\mu| < 2.4.
\end{equation}
The electrons must satisfy
\begin{equation}
|p_{Te^+}| + |p_{Te^-}| > 44~{\rm GeV},\quad |\eta_e| <1.44.
\end{equation}
so that a multivariate discriminator can be used to separate $\gamma^*\rightarrow e^+e^- $ from jets or single electrons \cite{Khachatryan:2015lga}.\footnote{CMS trained a discriminator to identify electron pairs, which they claim to have an efficiency around $40\%$. We did not include this treatment in our simulations due to the lack of details on the discriminator. But it would not change our conclusion even if such an efficiency is included.}
The photon must satisfy the following acceptance and be well-separated from leptons
\begin{equation}
p_T^\gamma > 0.3 M_{\ell\ell\gamma},\quad |\eta_\gamma| < 1.44,\quad
\Delta R_{\gamma\ell} > 1. 
\end{equation}
We would like to point out that, given the well-predicted kinematical properties of a fully reconstructable decay of the Higgs boson, the analyses may be improved by further utilizing the signal kinematical features. One of striking features is the mono-chromatic nature of the photon as given in Eqs.~(\ref{eq:ez}) and ~(\ref{eq:eg}). 
We thus propose to tighten the Higgs mass cut in Eq.~(\ref{eq:mcut}) as much as experimentally feasible, then boost the system to the Higgs boson rest frame, and impose the following cuts
\begin{equation}
60< E_\gamma <63~{\rm GeV\ in\ the\ rest\ frame\ of\ }\ell\ell\gamma,
\label{new_cuts_1}
\end{equation}
Another alternative option is to tighten the transverse momentum cut on the photon,
\begin{equation}
p_T^\gamma > 55~{\rm GeV}.
\label{new_cuts_2}
\end{equation}
The comparison of different cuts are demonstrated in Table \ref{tab:lepton}, where the cross sections of signals and backgrounds, as well as the statistical significances are listed. The first row in each block in Table \ref{tab:lepton} is calculated using the CMS acceptance cuts in Eqs.~(\ref{cut:z1})$-$(\ref{cut:z3}), and therefore serves as the reference to illustrate possible improvements by imposing our addtional cuts based on kinematical features.
Due to the stronger enhancement near the photon pole $\gamma^* \rightarrow e^+e^-$, one would be able to reach a $4.5\sigma/14\sigma$ sensitivity for the channel $h \rightarrow e^+e^-\gamma$ at the LHC with an integrated luminosity $0.3~{\rm ab}^{-1}/3~{\rm ab}^{-1}$, and a $3.1\sigma/9.9\sigma$ for the channel $h \rightarrow \mu^+\mu^-\gamma$. It is interesting to compare our results for the radiative decay $h \to \mu^+\mu^-\gamma$ with the ATLAS projection \cite{ATL-PHYS-PUB-2013-014} for the direct decay $h\to \mu^+\mu^-$ with the sensitivity reach of $2.3\sigma/7.0\sigma$ for $0.3~{\rm ab}^{-1}/3~{\rm ab}^{-1}$. 
Similar results have also been obtained by the CMS collaboration \cite{CMS:2013xfa}.


\subsection{$h\rightarrow Z\gamma\rightarrow\ell^+\ell^-\gamma$}
To select the signal events near the $Z$-pole from the Higgs decay and effectively suppress the backgrounds, we first follow the CMS analysis \cite{Chatrchyan:2013vaa} and require the invariant masses of the final state particles to be 
\begin{equation}
M_{\ell\ell} > 50~{\rm GeV}, \quad 120~{\rm GeV}<M_{\ell\ell\gamma}<130~{\rm GeV}.
\label{cut:z1}
\end{equation}
The leading (sub-leading) lepton must satisfy the acceptance of the transverse momentum and pseudo-rapidity 
\begin{equation}
p_T^\ell > 20~(10)~{\rm GeV},\quad |\eta_\mu| < 2.5,\quad |\eta_e|<2.4 .
\label{cut:z2}
\end{equation}
The photon must satisfy the following acceptance and be well-separated from leptons\footnote{We also impose $1.44 < |\eta| < 1.57$ to simulate the CMS barrel-end cap transition region. Additional cuts from CMS $p_T^\gamma > (15/110)M_{\ell\ell\gamma}$ and $M_{\ell\ell\gamma}+M_{\ell\ell} > 185~{\rm GeV}$ have been also adopted.}
\begin{equation}
p_T^\gamma > 15~{\rm GeV},\quad |\eta_\gamma| < 2.5,\quad \Delta R_{\gamma\ell} > 0.4. 
\label{cut:z3}
\end{equation} 
Similarily to the $h\rightarrow \gamma^*\gamma\rightarrow\ell^+\ell^-\gamma$ study, we again propose to tighten the energy and momentum cuts 
\begin{align}
27< & E_\gamma < 33~{\rm GeV\ in\ the\ rest\ frame\ of\ }\ell\ell\gamma, \label{new_cuts_3}\\
&p_{T}^\gamma>25~{\rm GeV},
\label{new_cuts_4}
\end{align}
as listed in Table \ref{tab:lepton}. Although the tight cuts do not improve the statistical significance significantly for these channels, the signal-to-background ratios are improved by about a factor of two, reaching a $1\%$ level. This would help to keep potential systematic errors in better control.
Unlike the $\gamma$-pole feature discussed above, there is no appreciable difference between $e^+e^-$ and $\mu^+\mu^-$ channels. One would be able to reach a $1.7\sigma/5.5\sigma$ sensitivity at the LHC with an integrated luminosity $0.3~{\rm ab}^{-1}/3~{\rm ab}^{-1}$. Although weaker signals than the $\gamma^*\gamma$ channels above, these will significantly improve the overall observability for $h \rightarrow\ell^+\ell^-\gamma$ if the analyses can be combined.

\begin{table}[t]
\begin{center}
\renewcommand{\arraystretch}{1.5}
\begin{tabular}{|c|c|c|c|}
\hline
Channel & Signal & Background & Statistical Significance \\
              & [fb] & [fb] & with 0.3 (3) ${\rm ab}^{-1}$ luminosity \\
\hline\hline
$pp\rightarrow \gamma^*\gamma\rightarrow \mu^+\mu^-\gamma$ & 0.69 & 23.5 & 2.47 (7.79) \\
$60 < E_{\gamma} < 63~{\rm GeV}$  & 0.69 & 14.6 & 3.13 (9.89) \\
$p_{T\gamma} > 55~{\rm GeV}$  & 0.46 & 11.8 & 2.32 (7.33) \\
\hline
$pp\rightarrow \gamma^*\gamma\rightarrow e^+e^-\gamma$ & 1.06 & 27.0 & 3.53 (11.2) \\
$60 < E_{\gamma} < 63~{\rm GeV}$  & 1.06 & 17.0 & 4.45 (14.1) \\
$p_{T\gamma} > 55~{\rm GeV}$  & 0.79 & 17.6 & 3.26 (10.3) \\
\hline  \hline
$pp\rightarrow Z\gamma\rightarrow \mu^+\mu^-\gamma$ & 1.40 & 214 & 1.66 (5.24) \\
$27 < E_{\gamma} < 33~{\rm GeV}$  & 1.10 & 121 & 1.73 (5.48) \\
$p_{T\gamma} > 25~{\rm GeV}$  & 0.91 & 95.9 & 1.61 (5.09) \\
\hline
$pp\rightarrow Z\gamma\rightarrow e^+e^-\gamma$ & 1.38 & 224 & 1.60 (5.05) \\
$27 < E_{\gamma} < 33~{\rm GeV}$  & 1.13 & 126 & 1.74 (5.51) \\
$p_{T\gamma} > 25~{\rm GeV}$  & 0.91 & 100 & 1.58 (4.98) \\
\hline
\end{tabular}
\end{center}
\caption{The cross sections of signals and backgrounds, and the statistical significances of $pp\rightarrow V\gamma \rightarrow \ell^+\ell^-\gamma$, $V=Z, \gamma^*$.}
\label{tab:lepton}
\end{table}


\subsection{$h\rightarrow J/\psi\ \gamma\rightarrow\ell^+\ell^-\gamma$}
\label{sec:Jpsi}
With respect to another similar final state from the Higgs boson decay, a comparative remark is in order.
It has been pointed out that the Higgs rare decay to a photon associated with a heavy vector meson $J/\psi$ may provide the direct access to the charm-Yukawa coupling via the clean leptonic decay channels \cite{Bodwin:2013gca}. The branching fraction in the SM is predicted \cite{Bodwin:2014bpa, Koenig:2015pha, Bodwin:2016edd} to be
\begin{equation}
{\rm BR_{SM}}(h\rightarrow J/\psi\ \gamma) = 2.79\times10^{-6}\ \ {\rm and}\ \  
{\rm BR_{SM}}(h\rightarrow J/\psi\ \gamma \to \mu^+ \mu^- \gamma) = 2.3 \times10^{-7},
\label{eq:brc}
\end{equation}
which is very small. Furthermore, the ``direct contribution" involving the charm-Yukawa coupling is much smaller than that from the ``indirect contribution'' via $\gamma^*\to J/\psi$ \cite{Bodwin:2014bpa}, making the probe to the charm-Yukawa coupling in this channel extremely challenging.

Nevertheless, for comparison, this result has been marked in Figs.~\ref{plot:decay_Mmm} and \ref{plot:decay_Mee}, in units of keV and without the photon acceptance cuts. The superb muon pair mass resolution of the order 100 MeV would be needed in order to have a chance to dig out the weak signal from the continuum $h\to \gamma^*\gamma \to \ell^+\ell^- \gamma$ events, on top of the other SM background sources. We propose to start with the larger event samples of $\ell^+\ell^-\gamma$ as discussed in the last two sections, relax the $J/\psi$-specific cuts in the hope for an early observation of the $h\to \ell^+\ell^-\gamma$ signal, and then to extend the search 
to scrutinize the potential excess from $J/\psi\to \ell^+\ell^-$.

Dedicated searches for this decay channel have been performed by ATLAS \cite{Aad:2015sda} and CMS \cite{Khachatryan:2015lga}. With $20~{\rm fb}^{-1}$ luminosity, both ATLAS and CMS set a bound of ${\rm BR}(h\rightarrow J/\psi\ \gamma) < 1.5\times10^{-3}$ under the assumption of SM Higgs production. 
If the beyond-the-Standard-Model (BSM) physics only enhances the charm-Yukawa coupling by a factor of $\kappa_c$,
\begin{equation}
y_c^{\rm BSM} = \kappa_c y_c^{\rm SM},
\label{kappa_def}
\end{equation}
then this experimental bound can be translated into a loose bound on $\kappa_c \lesssim 220$ \cite{Perez:2015aoa}. 
With 3 ab$^{-1}$ luminosity at the HL-LHC, the expected upper limit to 
${\rm BR_{SM}}(h\rightarrow J/\psi\ \gamma)$ is about 15 times the SM value \cite{ATL-PHYS-PUB-2015-043}, which corresponds to a upper bound of about $\kappa_c \lesssim 50$. 


\subsection{$h \rightarrow \tau^+\tau^-\gamma$}
\label{sec:tau}
Besides the clean $e^+ e^- \gamma,\ \mu^+\mu^- \gamma$ final states, the $\tau^+\tau^-\gamma$ channel is  also of considerable interests from the observational point of view. The direct decay $h\rightarrow\tau^+\tau^+$ has been observed in the LHC experiments mainly via the vector-boson-fusion production mechanism \cite{Aad:2015vsa,Chatrchyan:2014nva}. The radiative decay channel $h\rightarrow\tau^+\tau^+\gamma$ may be searched for via the leading production channel of gluon fusion. To compare the rates, we have at the 14 TeV LHC, 
\begin{eqnarray}
\label{eq:vbf}
&& \sigma(WW,ZZ\to h \to \tau^+\tau^-) = (4.2~{\rm pb}) \times (6.3\%) \approx 260~{\rm fb}; \\
&& \sigma(gg\to h \to \tau^+\tau^- \gamma) = (49~{\rm pb}) \times (0.1\%) \approx 50~{\rm fb}. 
\label{eq:gg}
\end{eqnarray}
Thus, it is quite conceivable to observe this radiative decay mode in the future searches. The kinematical features of this decay will be rather different from those presented in the last sections due to the dominance of the QED radiation. Because of the complexity of the tau decay final states, the signal observation and the background suppression will need to be carefully analyzed \cite{Galon:2016ngp}. 
We will leave this to a future analysis.


\section{LHC Search for $c\bar c\gamma$ and the Charm-Yukawa Coupling}
\label{sec:LHCcc}

It is crucially important to search for the decay $h\to c\bar c$, since it is the largest mode for the Higgs boson to couple to the second generation of fermions, which would be sensitive to the physics beyond the Standard Model. It has been pointed out that the charm-Yukawa coupling could be significantly modified in various BSM models \cite{Giudice:2008uua, Bauer:2015fxa, Harnik:2012pb, Delaunay:2013iia, Delaunay:2013pwa, Blanke:2013zxo, Kagan:2009bn, Dery:2013aba, DaRold:2012sz, Dery:2014kxa, Bishara:2015cha}.
%
Given the difficulty as seen above in searching for $h\to J/\psi\ \gamma \to \ell^+\ell^- \gamma$, other methods have also been explored to probe the charm-Yukawa coupling \cite{Delaunay:2013pja, Zhou:2015wra, Perez:2015lra, Brivio:2015fxa, Bishara:2016jga, Yu:2016rvv, Carpenter:2016mwd}. In this section, we discuss the possibility of constraining the charm-Yukawa coupling using the open-flavor channel $pp\rightarrow c\bar{c}\gamma$, which has a much larger branching fraction about $4\times 10^{-4}$, as seen in Fig.~\ref{fig:br}. The additional photon radiation may serve as the trigger \footnote{A 20~GeV cut on the transverse momentum of the photon would be too low as a trigger threshold for a single photon. However, it is conceivable to design an improved trigger strategy including the $c\bar c$ final state. Realistic trigger threshold and background rate are under careful investigation now \cite{paper:h2cca_trigger}.} and is in favor of picking out the $c\bar c$ events over $b\bar b$ due to the larger charm electric charge. 

\begin{table}[t]
\begin{center}
\renewcommand{\arraystretch}{1.5}
\begin{tabular}{|c|c|c|c|}
\hline
Operating Point & $\epsilon_c$ & $\epsilon_b$ & $\epsilon_j$ \\
\hline\hline
I & 20\%  & 10\%  & 1\%  \\
II & 30\%  & 20\%  & 3\%  \\
III & 45\%  & 50\%  & 10\%  \\
\hline
\end{tabular}
\end{center}
\caption{Representative operating points for the $c$-tagging efficiency ($\epsilon_c$), $b$ and light jets contamination rates ($\epsilon_b$ and $\epsilon_j$).}
\label{tab:c-tag}
\end{table}

The signal events are characterized by a high-$p_T$ photon recoiling against a pair of charm-jets. To identify such events, an efficient charm-tagging technique is required. Although currently there is no dedicated charm-tagging being implemented at the LHC, the discrimination of a $c$-jet from a $b$-jet has been studied and used in the calibration of the $b$-tagging efficiency \cite{ATLAS:2012mma, ATLAS:2014pla}. ATLAS also proposed a $c$-tagging algorithm \cite{ATLAS:2015ctag} based on the neural network that could achieve about 20\% (90\%) tagging efficiency with a medium (loose) cut criteria in the search for $pp\rightarrow \tilde{t}\tilde{t}^*\rightarrow (c \tilde{\chi}^0_1)(c \tilde{\chi}^0_1)$. In the current study, we choose three representative operating points listed in Table \ref{tab:c-tag}, for the $c$-tagging efficiency $\epsilon_c$, and $b$ and light jets contamination rates, $\epsilon_b$ and $\epsilon_j$, respectively. When increasing the $c$-tagging efficiency from I to III, we must accept higher contaminations from a heavier quark and light jets.

\begin{table}[t]
	\begin{center}
		\renewcommand{\arraystretch}{1.5}
		\begin{tabular}{|c|c|c|c|c|c|}
			\hline
			Luminosity                 &Operating  & Signal     &  Signal       &  Signal            & Background             \\
			&Point          &(Total)         & (QED)       &(EW$+\gamma$)          &                                 \\
			\hline\hline
			&I                 &  683        & 252            & 431                 & $3.84\times10^{7}$    \\
			$3000~{\rm fb}^{-1}$ &II                &  1537      &567             &970         	     & $1.25\times10^{8}$    \\
			&III               &  3459	  &1275           & 2184	              & $6.51\times10^{8}$    \\
			\hline
		\end{tabular}
	\end{center}
\caption{Numbers of events for the signals and backgrounds with the three $c$-tag operating points for an integrated luminosity of 3000 fb$^{-1}$.}
\label{tab:c-Xsec}
\end{table}

The dominant background is the QCD di-jet plus a direct photon production, with the jets to be mis-tagged as $c$-jets. Another major background is the QCD 3-jet production, leading to two mis-tagged $c$-jets associated with a fake photon radiation. Following an ATLAS analysis \cite{ATL-PHYS-PUB-2011-007}, we take the photon fake rate from a light-quark jet and from a gluon jet to be 
\begin{equation}
\epsilon_{q\to \gamma} = 0.06\% ,\quad \epsilon_{g\to \gamma} = 0.006\% ,
\end{equation}
respectively.
We note that the fake photon contamination contributes about $(10-30)\%$ to the total background. Another potentially large background is from jet fragmentation into a real photon. We assume that the stringent photon isolation requirement will be sufficient to suppressed this QCD background, as pointed out in the prompt photon studies \cite{Catani:2002ny}. In our simulations, we require that both the $c$-jets and the photon be hard and well-isolated in the central region
\begin{equation}
p_{T}~>~20~{\rm GeV},\;\; |\eta| < 2.5,\;\; {\rm and} \;\;\Delta R > 0.4.
\end{equation}
The ultimate sensitivity for the signal $h\to c\bar c \gamma$ depends on the invariant mass reconstruction $M_{jj\gamma} = m_h$, and thus the energy resolution of the charm-jets. 
In this study, we assume that the Higgs resonance peak can be reconstructed within $20\%$ and thus we require
\begin{equation}
100~{\rm GeV} < M_{jj\gamma} < 150~{\rm GeV}.
\end{equation}
Tightening this mass cut would linearly improve the signal-to-background ratio. We also apply $ p_{T}^{\rm max} > 40 ~{\rm GeV}$ to further increase the signal-to-background ratio $S/B$. With these cuts applied, the background rate at the HL-LHC would be controlled below 1 kHz, within the detector's trigger ability. A fully implementable trigger scheme and the cut optimization are under investigation.
After the above cuts applied, we list the numbers of events in Table \ref{tab:c-Xsec} for an integrated luminosity of 3000 fb$^{-1}$. We note that, within the SM, the signal events from the QED radiation and the EW$+\gamma$ processes are comparable, unlike the situation in $h\rightarrow J/\psi\ \gamma$ where the dominant contribution is from the ``indirect contribution'' via $\gamma^* \rightarrow J/\psi$. Unfortunately, with the Standard Model predictions for the signal and backgrounds being $S/B < 10^{-4}$, it would not be promising to observe this channel at the HL-LHC.

\begin{table}
  \center
  \begin{tabular}{|c|c|}
	\hline
	Method & $\kappa_c$ upper limit projection\\
	             & at HL-LHC ($\rm 3~ab^{-1}$) \\
	\hline\hline
	$h\to c\bar c \gamma$ (this work) & 6.3 \\
	\hline
	$h\to c\bar c+$fit \cite{Perez:2015lra} & 2.5 	\\
	\hline
	$h+c$ production \cite{Brivio:2015fxa}  & 2.6 	\\
	\hline
	Higgs kinematics \cite{Bishara:2016jga} & 4.2\\
	\hline
	$h\to J/\psi \gamma$ \cite{ATL-PHYS-PUB-2015-043} & 50\\
	\hline
\end{tabular}
  \caption{Projected sensitivities for probing the $hc \bar c$ Yukawa coupling $\kappa_c=y_c^{\rm BSM}/y_c^{\rm SM}$ at the HL-LHC with various methods. }
  \label{tab:kappac}
\end{table}

If the BSM physics significantly modifies the charm-Yukawa coupling as parameterized in Eq.~(\ref{kappa_def}), then the QED radiation will be scaled by a factor of $\kappa_c^2$. In principle, such a deviation would also change to rate through the Higgs total width. However, since the SM branching fraction is of $\mathcal{O}(10^{-4})$, we approximate the Higgs total width to be unchanged. Although both the QED radiation and EW$+\gamma$ processes contribute to the signal, it would be dominated by the QED radiation if the charm-Yukawa coupling significantly deviates from the SM value. Therefore, considering only the statistical significance by the Gaussian standard deviation 
\begin{equation}
\sigma_{\rm SD} = \frac{N_{\rm S}^{\rm BSM}}{\sqrt{N_{\rm B}}} \simeq \frac{ \kappa_c^2\ N_{\rm S}^{\rm QED} }{\sqrt{N_{\rm B}}},
\end{equation}
the $2\sigma$-bounds on the charm-Yukawa coupling are obtained as 
\begin{equation}
\kappa_c < 12.5\;(7.0),\;\; 11.1\;(6.3),\;\;11.2\;(6.3).
\end{equation}
for operating points I, II, III with a luminosity of $3000~{\rm fb^{-1}}$. Those results with the Higgs radiative decay, although still rather weak, could be comparable to the recent studies on the charm-Yukawa coupling \cite{Perez:2015lra,Brivio:2015fxa,Bishara:2016jga,Carpenter:2016mwd} and seem to be more advantageous to $h\to J/\psi\  \gamma$. We have complied the existing results in Table \ref{tab:kappac}. The first three methods listed here rely on different production mechanisms and certain charm-tagging techniques with various assumptions of $c$-tagging efficiencies.\footnote{The authors in \cite{Perez:2015lra} used an integrated luminosity of $2\times 3000~\rm fb^{-1}$ (combining both the ATLAS and CMS data), and the tagging efficiencies $\epsilon_c =0.5$, $\epsilon_b=0.2$, and $\epsilon_j=0.005$; while the authors in  \cite{Brivio:2015fxa}  adopted the tagging efficiencies $\epsilon_c =0.4$, $\epsilon_b=0.3$, and $\epsilon_j=0.01$. If using their choices for our analysis, we would have gotten a slightly stronger bound with $\kappa_c < 4.2$ and 4.9, respectively.}
Nevertheless, they tend to have better performances than the $h\to J/\psi\ \gamma$ channel, mainly because of the larger signal rates for the open $c$-flavor production. Those channels should thus be complementary in the future explorations.

Before closing this section, we would like to comment on the $h\rightarrow b\bar{b}\gamma$ channel at the LHC. The current measurement on the $h\rightarrow b\bar{b}$ channel is mainly through $q\bar q\to Vh$ production and already of about $3\sigma$ significance with current data at the LHC \cite{Aad:2014xzb, Chatrchyan:2013zna}. Although dominated by the QED radiation, the $h\rightarrow b\bar{b}\gamma$ channel is scaled down further by the bottom-quark electric charge squared, a factor of 4, compared to $h\rightarrow c\bar{c}\gamma$. As listed in Table~\ref{tab_width}, the braching fraction of $h\rightarrow b\bar{b}$ with $E_\gamma > 15~{\rm GeV}$ is about 500 times less than that of $h\rightarrow b\bar{b}$. Therefore, it would be less promising for the $h\rightarrow b\bar{b}\gamma$ channel to compete with $h\rightarrow b\bar{b}$, in contrast to our analysis above for $h\rightarrow c\bar{c}\gamma$.


\section{Summary}
\label{sec:sum}
With a large data sample of the Higgs boson being accumulated at the LHC or anticipated at the HL-LHC, it is strongly motivated to search for rare decays of the Higgs boson to test the Higgs sector in the SM and to seek for hints of BSM physics. In this work, we studied the Higgs rare decay channels $h\rightarrow f\bar{f}\gamma$ where $f = \tau,\mu,e$ and $b,c$ and their observability at the LHC. Our results can be summarized as follows.
\begin{itemize}
\item 
This radiative decay channel receives contributions from QED corrections to the Yukawa interactions at $\mathcal{O}(y_f^2\alpha)$ and EW$+\gamma$ processes at $\mathcal{O}(y_t^2\alpha^3,\alpha^4)$, as we discussed in Sec.~\ref{sec-QED} and \ref{sec-EW}. The QED corrections constitute about $Q_f^2 \times \mathcal{O}(1\%)$ to the partial widths of fermionic Higgs decays  in particular through the running mass, and therefore should be taken into account for future precision Higgs physics. The difference between the QED resummed running mass in Eq.~(\ref{eq:qcd_running_mass}) and its $\mathcal{O}(\alpha)$ approximation only contributes to the NLO QED corrections at percentage level, due to the weakly-coupled nature of QED, as shown in Table~\ref{tab_mass}.
\item 
As showed in Sec.~\ref{section:decay_width}, the contributions from the Yukawa corrections (Fig.~\ref{feyn:qed}) and the EW$+\gamma$ contributions (Fig.~\ref{feyn:ew}) exhibit quite different patterns for different fermions in the final state: While they are comparable for $c\bar c \gamma$, the Yukawa corrections dominate for $b\bar b \gamma,\ \tau^+\tau^-\gamma$. The EW$+\gamma$ loops overwhelm for $\mu^+\mu^-\gamma,\ e^+ e^-\gamma$, which results in the branching fractions of the order $\mathcal{O}(10^{-4})$ despite their tiny Yukawa couplings (see Fig.~\ref{fig:br}). The main contributions in the EW$+\gamma$ loops are around the $Z$-pole, as well as the $\gamma$-pole near $m_{\gamma^*}\approx 2m_f$. The kinematic distributions, especially 
the photon energy distributions in Fig.~\ref{plot:decay_Ea} and the invariant mass distributions in Fig.~\ref{plot:decay_Mff} are quite informative to reveal the underlying decay mechanisms, and to guide the experimental searches. 
\item
As the $e^+e^-\gamma$ and $\mu^+\mu^-\gamma$ channels exhibit the violation of the Yukawa scaling, we studied their observability at the LHC in Sec.~\ref{sec:LHCll}, taking into account the signal characteristics and the SM background. We proposed new cuts based on the kinematical features in Eqs.~(\ref{new_cuts_1}),~(\ref{new_cuts_2}),~(\ref{new_cuts_3}),~and~(\ref{new_cuts_4}) in addition to the selection cuts by CMS. For $pp\rightarrow \gamma\gamma^*\rightarrow \ell^+\ell^-\gamma$ channels, the statistical significances and the siginal-to-background ratios are improved by about 25\% and 60\%, respectively. For $pp\rightarrow Z\gamma\rightarrow \ell^+\ell^-\gamma$ channels, the siginal-to-background ratios are enhanced by about 80\% while the statistical significances stay about the same.
We conclude that, with an integrated luminosity 0.3 ab$^{-1}$/3 ab$^{-1}$, 
the channels $h \rightarrow \gamma^*\gamma\rightarrow e^+e^-\gamma\ (\mu^+\mu^-\gamma)$ should be observable at the level of $4.5\sigma/14\sigma\ (3.1\sigma/9.9\sigma)$, and the channels $h \rightarrow Z\gamma\rightarrow e^+e^-\gamma,\ \mu^+\mu^-\gamma$ should be observable at the level of $1.5\sigma/5.5\sigma$. The sensitivity could be comparable to the direct search of the two-body decay $h\to \mu^+\mu^-$. 
\item
The decay $h \to J/\psi\ \gamma \to \ell^+ \ell^- \gamma$ has the same final state but much smaller rate. The searches for the above channels will serve as the necessary early discovery and will shed light on the potential observation for $h\to J/\psi\ \gamma$. 
\item
In Sec.~\ref{sec:tau}, we pointed out a potentially observable decay $h \to \tau^+ \tau^- \gamma$. We proposed the search via the leading production mechanism from gluon fusion with the help of the additional photon. 
\item
In Sec.~\ref{sec:LHCcc}, we proposed to probe the charm-Yukawa coupling via the decay channel $h\rightarrow c\bar{c}\gamma$. With the help of future $c$-tagging techniques, we demonstrated that the charm-Yukawa coupling $y_c$ can be bounded as $y_c^{\rm BSM} \lesssim 6y_c^{\rm SM}$ at $2\sigma$ level at the HL-LHC. We find it potentially comparable to the other related studies in the literature, and better than the $J/\psi\ \gamma$ channel in constraining the charm-Yukawa coupling. A more comprehensive analysis with realistic simulations is under way.
\end{itemize}

\acknowledgments We would like to thank Kaoru Hagiwara and Ayres Freitas for helpful discussions. This work is supported in part by the Department of Energy under Grant No.~DE-FG02-95ER40896 and in part by PITT PACC. XW was also supported in part by a PITT PACC Predoctoral Fellowship from School of Arts and Sciences at the University of Pittsburgh. 

\bibliographystyle{JHEP}
\bibliography{h2ffa_refs} 
\end{document}